\documentclass[12pt]{article} 
\usepackage[utf8]{inputenc} 
\usepackage{lmodern}  
\usepackage{amsmath} 
\usepackage{amsfonts}  
\usepackage{amssymb}  
\usepackage{todonotes} 
\usepackage{csquotes}
\usepackage[UKenglish]{babel} 
\usepackage{subcaption}
\usepackage{booktabs}
\usepackage{authblk}

\usepackage[pdfauthor="L Glaser",pdftitle="Computational\ explorations\ of\ a\ deformed\ fuzzy\ sphere"]{hyperref}

\usepackage[sorting=none,maxnames=5,giveninits=true,doi=true,url=true]{biblatex}
\DeclareCiteCommand{\fullcite}
  {\usebibmacro{prenote}}
  {\clearfield{url}%
    \clearfield{doi}
    \clearfield{issn}
    \clearfield{title}
   \clearfield{pages}%
   \clearfield{pagetotal}%
   \clearfield{edition}%
   \clearfield{labelyear}%
   \usedriver
     {\DeclareNameAlias{sortname}{default}}
     {\thefield{entrytype}}}
  {\multicitedelim}
  {\usebibmacro{postnote}}

\newcommand{\mc}[1]{\mathcal{#1}}

\newcommand{\Mn}{M_n(\mathbb{C})}

\title{Computational explorations of a deformed fuzzy sphere}
\author[1]{L Glaser \thanks{L.glaser@univie.ac.at}}
\affil[1]{University of Vienna, Faculty of Physics, Boltzmanngasse 5, 1090 Vienna, Austria}

\begin{document}
\maketitle
\begin{abstract}
  This work examines the deformed fuzzy sphere, as an example of a fuzzy space that can be described through a spectral triple, using computer visualizations.
  We first explore this geometry using an analytic expression for the eigenvalues to examine the spectral dimension and volume of the geometry.
  In the second part of the paper we extend the code from~\fullcite{Glaser_Stern_2021}, in which the truncated sphere was visualized through localized states.
  This generalization allows us to examine finite spectral triples.
  In particular, we apply this code to the deformed fuzzy sphere as a first step in the more ambitious program of using it to examine arbitrary finite spectral triples, like those generated from random fuzzy spaces, as show in~\fullcite{Barrett_Glaser_2016}.
\end{abstract}
\section{Introduction}
Quantizing gravity is one of the most difficult open questions remaining in fundamental physics.
One way to approach this problem is to closer examine the mathematical structure underlying our theories, to rewrite them in a more elegant way.
Recasting the differential geometry of a manifold as the algebraic data of a spectral triple is one such way.
Writing geometry in an algebraic language like this is promising since many of quantum theory's unintuitive results arise from its underlying algebraic structure.

A spectral triple consists of an algebra, a Hilbert space and a Dirac operator $(\mc{A},\mc{H},D)$. 
Any compact Riemannian manifold can be described as a commutative spectral triple which fulfils a number of axioms, as proposed by Connes in~\cite{Connes_1994}.
He showed in particular that the metric information of the manifold can be recovered using the distance between pure states $s_1,s_2$ of the algebra, defined as
\begin{align}\label{eq:ConnesDist}
  d(s_1,s_2)= \sup_{a\in \mc{A}} \left\{ |s_1(a)-s_2(a)| \big |[D,a]| \leq1 \right\} \;.
\end{align}
In this article we will be studying real finite spectral triples as defined by John Barrett in~\cite{Barrett_2015}.
A real spectral triple $(s,\mc{A},\mc{H},D,J,\Gamma)$ arises when describing a spin manifold.
In addition to the algebra, Hilbert space and Dirac operator it has a KO-dimension $s$, a real structure $J$  and a chirality $\Gamma$.
The spectral triples we examine are a subclass called matrix geometries in~\cite{Barrett_2015}.
For these the algebra consists of $n$ by $n$ matrices $\mc{A}=\Mn$, and the Hilbert space is a product space $\mc{H}= V \otimes \Mn$, with $V$ a $(p,q)$-Clifford module.
The $(p,q)$-Clifford module is generated by $q$ $\gamma$-matrices with $\gamma_i^2=-1$ and $p$ $\gamma$-matrices with $\gamma_i^2=1$.
The KO-dimension is then given as $s=q-p \, \mod \, 8$.
The algebra acts on the Hilbert space through a representation $\rho(a)$ with $\rho(a)(v\otimes m)= v \otimes a\cdot m$ where $ m,a \in \Mn$ with $a \cdot m$ the usual matrix product and $v \in V$.
The real structure $J$ connects the left action of the algebra on the Hilbert space to the right action as $\rho_l(a) = J \rho_r(a) J^{-1}$.
The Dirac operator on such a spectral triple has to fulfil a number of conditions; it should be self adjoint, commute or anti-commute with $J$ and $\Gamma$ depending on the KO-dimension, and satisfy the first order condition
\begin{align*}
[[D,\rho_r(a)],\rho_l(b)]&= 0 \quad \forall \quad  a,b \in \mc{A} \;.
\end{align*}
Taking all of these conditions on the Dirac operator into account leads to a restricted parametrization of the Dirac operator, purely through a set of hermitian and anti-hermitian matrices in $\Mn$
\begin{align}\label{eq:Dgeneral}
  D(v \otimes m)= \sum_i \omega_i v \otimes (K_i m + \epsilon' K_i^*) 
\end{align}
where $\omega_i$ are products of $\gamma$ matrices and $K_i$ are matrices that are hermitian or anti-hermitian respectively, depending on the sign $\epsilon'$ and whether the product $\omega_i$ is hermitian or anti-hermitian (more detail on this is given in~\cite{Barrett_2015}).

This parametrization becomes immediately obvious if we look at the well known Grosse-Presn\v ajder Dirac operator for the fuzzy sphere
\begin{align}\label{eq:GP}
D_{GP} (v\otimes m)= v\otimes m + \sum_{j<k} \sigma^j \sigma^k v \otimes [L_{jk},m] \;,
\end{align}
where $\sigma^j$ are the Pauli matrices and $L_{jk}$ are the generators of the Lie algebra $so(3)$ in an $n$-dimensional irreducible representation on $\mc{H}$. 
This operator acts on a Hilbert space with a $(0,3)$-Clifford module, leading to KO-dimension $s=3$, while the sphere has KO-dimension $s=2$.
It also leads to a non-symmetric spectrum, since the largest eigenvalue of the operator has no corresponding negative eigenvalue.
These points led Barrett to propose an alternative Dirac operator for the fuzzy sphere, acting on a $(1,3)$-Clifford module 
\begin{align}\label{eq:DS2}
D_{S^2} (v \otimes m)=  \gamma^0 v \otimes m + \sum_{j<k} \gamma^0 \gamma^j \gamma^k v \otimes [L_{jk},m]\;,
\end{align}
where now the $\gamma^j$ are $(1,3)$-Clifford matrices and ${(\gamma^0)}^2=1$.
This operator consists of two copies of the Grosse-Presn\v ajder operator, acting on the $\pm1$ subspaces of the matrix $\gamma^{0}$.

Parametrizing the Dirac operator through a set of matrices has opened the door to defining an ensemble of finite spectral triples~\cite{Barrett_Glaser_2016,Glaser_2017}.
The premise of this work was, that finite spectral triples as defined above are a generalization of discrete geometries, which certainly includes well known cases like the fuzzy sphere.
Thus fixing the algebra and Hilbert space, while varying the Dirac operator over all operators parametrized as in equation~\eqref{eq:Dgeneral} would create an ensemble of geometric and not so geometric spectral triples
\begin{align}
\mc{Z} = \int \mc{D}[D] e^{- \mc{S}(D)} \;.
\end{align}
Using the Haar measure on the space of matrices $K_i$ and defining some spectral action $\mc{S}(D)=Tr(f(D))$ does then allow us to define this path integral over finite spectral triples, as a multi matrix model
\begin{align}
 \mc{Z} = \int \mc{D}[K_i] e^{- \mc{S}(K_i)} \;.
\end{align}
This may be studied using computer simulations, as e.g.\ done in~\cite{Barrett_Glaser_2016,Barrett_Druce_Glaser_2019,Glaser_2017,MauroDArcangelo_2022} or using analytic techniques~\cite{Azarfar_Khalkhali_2019,Khalkhali_Pagliaroli_2020,Hessam_Khalkhali_Pagliaroli_2021,Pérez-Sánchez_2021,Khalkhali_Pagliaroli_2021,Hessam_Khalkhali_Pagliaroli_2023,Verhoeven_2023}.
The action used for these studies was a simple quartic action $\mc{S}(D)= Tr(D^4) + g_2 Tr(D^2)$, with a single coupling constant $g_2$.
Exploring a range of possible values of $(p,q)$ and searching for a phase transition in $g_2$ showed that even this simple model has a rich structure.
A closer look at spectral observables, like the dimension, then indicated that the model has the potential to dynamically favour geometric spectral triples, in particular at phase transition points~\cite{Barrett_Druce_Glaser_2019,Glaser_2017}.

These results lead to the question: are there any more tools to understand whether a finite spectral triple is associated to a geometry, and if it is, how the associated geometry looks?
A partial answer, from a slightly different context came in~\cite{Glaser_Stern_2021} where we developed an algorithm to visualize truncated spectral triples.
In the present article we will explore an extension of this code, towards visualizations of finite spectral triples, to try and recover their geometry.
The first step in this process is to visualize, and thus better understand, one particular class of finite spectral triples, which are the deformed fuzzy sphere.

This paper is split into five parts, starting with this introduction.
The second section then introduces the deformed fuzzy sphere, and studies its eigenvalues and some geometric properties which can be derived from these.
The third section describes the code used to visualize the deformed fuzzy sphere, with a particular focus on those parts of the code that were modified compared to past work.
In the fourth section we will analyse the visualizations of the deformed fuzzy sphere, and explain the images we find.
The last section consists of a conclusion and outlook, pointing towards possible further applications of this algorithm, and gaps within this work that might be usefully filled.

\section{The spectrum of a deformed fuzzy sphere}
\begin{figure}
  \begin{subfigure}{0.49\textwidth}
\includegraphics[width=\textwidth]{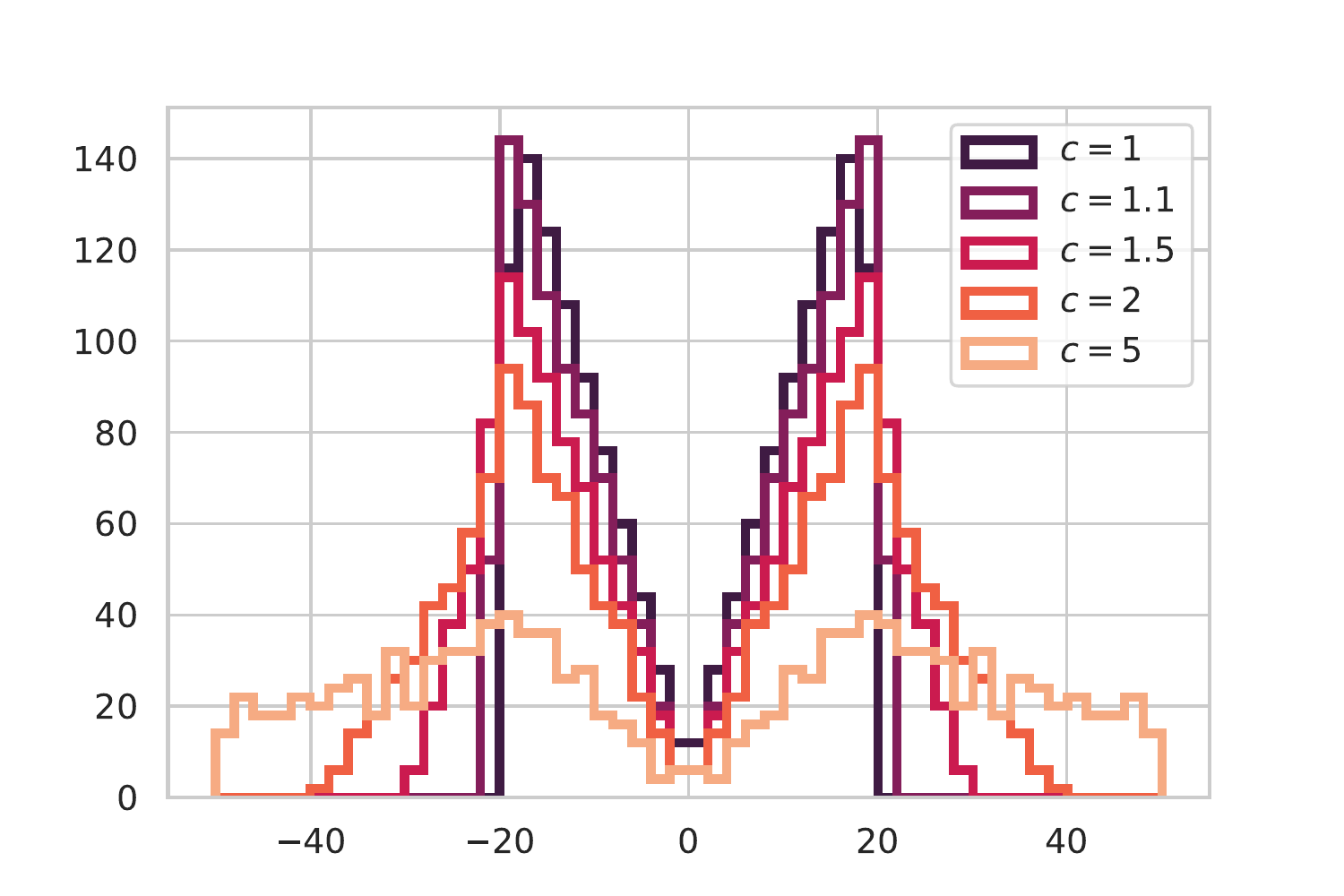}
\caption{$c\geq1$}
\end{subfigure}\hfill
\begin{subfigure}{0.49\textwidth}
\includegraphics[width=\textwidth]{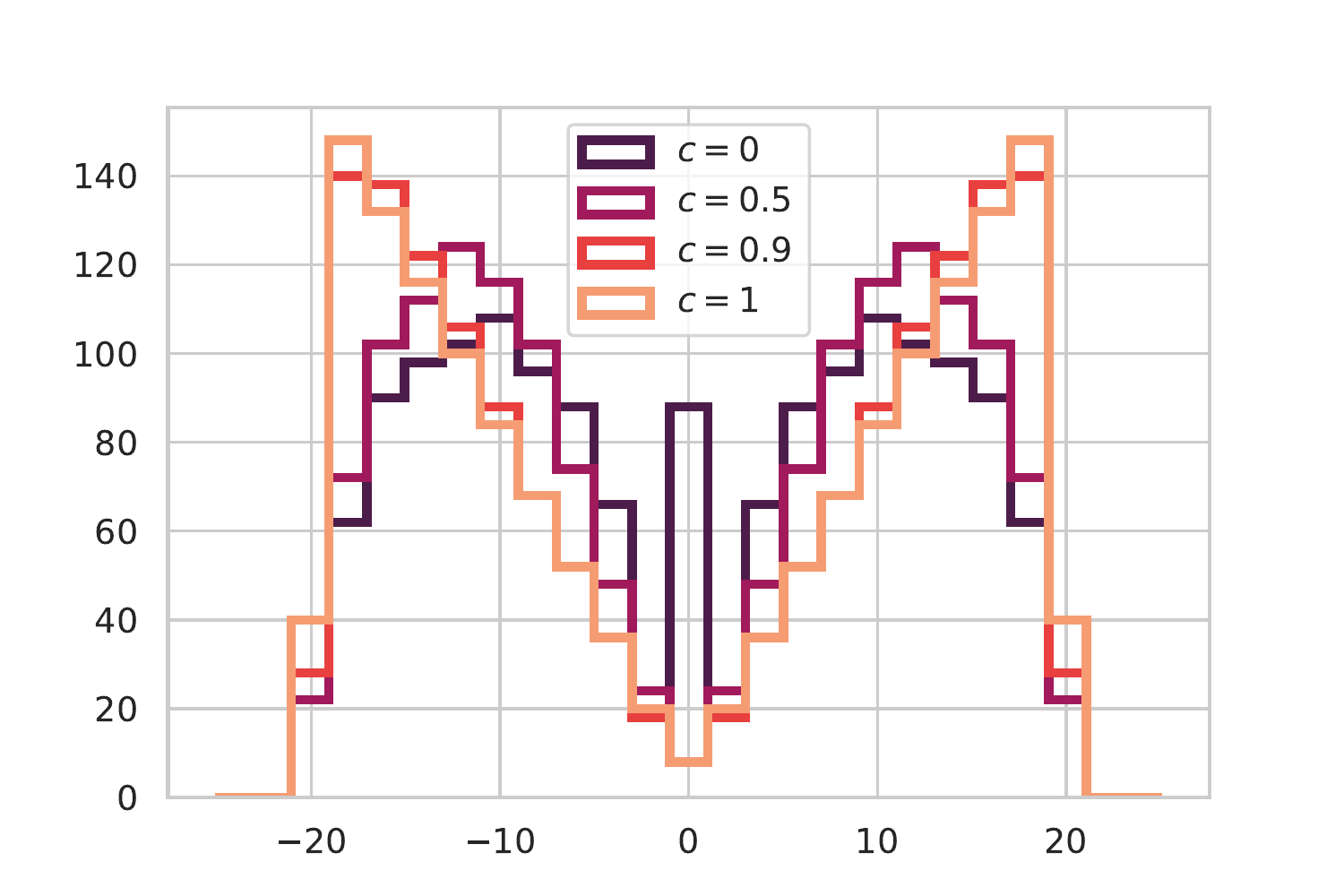}
\caption{$c\leq1$}
  \end{subfigure}
    \caption{Histogram of the eigenvalue spectrum for $n=20$ for a range of deformation parameters}\label{fig:spec_hist}
\end{figure}

An interesting modification of the fuzzy sphere, given in~\eqref{eq:DS2}, is the deformed fuzzy sphere, in which the Dirac operator is changed to be
\begin{align}
  D(v \otimes m) &=c_0 \gamma^0 v \otimes m + \sum_{j<k =1}^{3} c_{jk}\gamma^0 \gamma^j \gamma^k v \otimes [L_{j k}, m]
\end{align}
where $c_0, c_{jk}\in \mathbb{R}$ are deformation parameters.
If we restrict this most general extension to only introduce two parameters $a,c$
\begin{align}
  D(v \otimes m) &=a \gamma^0 v \otimes m + c \gamma^0 \gamma^1 \gamma^2 v \otimes [L_{1 2}, m]
  + \gamma^0 \gamma^1 \gamma^3 v \otimes [L_{1 3}, m]\\
  &+ \gamma^0 \gamma^2 \gamma^3 v \otimes [L_{2 3}, m] \;, \nonumber
\end{align}
we are able to find exact analytic results for the eigenvalues, very similar to how it is done for the ordinary fuzzy sphere.
The derivation of these eigenvalues, and a closer analytic study of the classical analogue of this operator will be published in future work~\cite{future}.
The analytic expression for the eigenvalues comes out as
\begin{align}
\lambda_{j,k }&=\pm \left( a - \frac{c}{2} + \sqrt{ j^2 + (c^2-1) k^2}  \right)\label{eq:evpos}\\
&\text{for } j = \frac{1}{2}, \frac{3}{2} \dots n-\frac{1}{2} & &k = \frac{1}{2}, \frac{3}{2}, \dots j \\
\lambda_{j,k }&=\pm \left( a - \frac{c}{2} - \sqrt{ {(j+1)}^2 + (c^2-1) k^2} \right) \label{eq:evneg}\\
&\text{for } j = \frac{1}{2}, \frac{3}{2} \dots n-\frac{3}{2} & &k = \frac{1}{2}, \frac{3}{2}, \dots j \;.
\end{align}
Each of these eigenvalues arises with multiplicity two, giving us a symmetric, real spectrum of eigenvalues for the deformed fuzzy sphere.
This spectrum, for $n=20$ is shown in Figure~\ref{fig:spec_hist}.

In the left-hand picture we look at $c\geq1$, there we see that as the deformation becomes stronger the maximal eigenvalue increases and the spectrum flattens out.
The central region still shows a linear rise, however this region becomes smaller as the deformation becomes stronger.

The right-hand image shows $0\leq c \leq 1$, and we can see that in this region the influence of the deformation is not as strong, although for $c=0$ eigenvalues of value $0$ appear. 

Using these results we can examine spectral observables for the deformed fuzzy sphere.
Two particularly interesting cases of this, which we will use in the visualization algorithm to decide how many coherent states to generate for each geometry, are the spectral dimension and the volume of the geometry.

Before we examine these we would like to understand what effect the deformation has on the shape of the geometry.
To study this  we examine the continuum equation for an ellipsoid 
\begin{align}
  r^2&= \frac{x_1^2}{\alpha^2}+\frac{x_2^2}{\beta^2}+\frac{x_3^2}{\gamma^2} \;,
\end{align}
and compare it to a continuum deformed Dirac operator
\begin{align}
D_{cont} &= (\sigma \cdot x) \sum_{i<j} c_{ij} \sigma^i \sigma^j \nabla_{X_{ij}}
\end{align}
where the vector fields for the sphere are $X_{ij}= x^i \partial_{x^j} - x^j\partial_{x^i}$.
To see how $\alpha, \beta,\gamma$ relate to the $c_{ij}$ we evaluate the continuum Dirac at special points, e.g.\ for $x=(0,0,1)$ the Dirac becomes
\begin{align}\label{eq:cont001}
D_{cont (0,0,1)}= c_{13} \sigma^1 \partial_{x^1} + c_{23} \sigma^2 \partial_{x^2} \;.
\end{align}  
For general $c_{ij}$ this operator is not symmetric along the two directions $x^1,x^2$.
However, if we rescaled a continuum ellipsoid as $x^1 \to x^{1}/\alpha,x^2 \to x^{2}/\beta,x^3 \to x^{3}/\gamma$ then the rescaling would lead to new coordinates in which the ellipsoid is a sphere.
So if we do this rescaling we find that $c_{13} \alpha$ and $c_{23} \beta$ should be the same, to lead to a symmetric operator in equation~\eqref{eq:cont001}.
This gives us a relation between $\alpha,\beta$ and $c_{12},c_{23}$.
Repeating this for the points $(1,0,0)$ and $(0,1,0)$ we find two more similar relations.
Using these we can, up to an overall prefactor, understand the relation between the axes lengths of the ellipse and the deformation parameters in our Dirac operator
\begin{align}\label{eq:defo_expl}
\alpha &= \frac{1}{c_{12} c_{13}} &\beta &= \frac{1}{c_{12}c_{23}} &\gamma &= \frac{1}{c_{13} c_{23}} \;.
\end{align}
This discussion has avoided considering the additional terms in the Dirac arising from the connection in the covariant derivative. 
However since the derivative terms definitely have to scale correctly this treatment is sufficient to fix the scaling.

\subsection{Spectral dimension}
Based on our understanding about the shape of the geometry we can then study spectral observables.
We begin with the spectral dimension, which was first studied in causal dynamical triangulations~\cite{Ambjørn_Jurkiewicz_Loll_2005}. 
It is a dimension measure that can resolve how the dimension of a geometry behaves with scale.
As a simple example, a $2$-sphere is $2$-dimensional if we are looking at length scales smaller than the radius of the sphere, but appears like a point if we are observing scales that are much larger than the radius.
A dimension spectrum like this is particularly interesting for quantum geometries, since their quantum structure can give rise to changes in dimension at small scales.

The prototypical example of this is in $4$ dimensional causal dynamical triangulations which find a UV scale dimension close to $2$, while at larger scales it recovers the expected $4$ dimensional behaviour before dropping to $0$ for low energies~\cite{Ambjørn_Jurkiewicz_Loll_2005}.
For fuzzy spaces the spectral dimension is defined from the eigenvalues of the Dirac operator $\{ \lambda \}$ as
\begin{align}
  d_s(t)&= 2t \; \langle \lambda^2 \rangle = 2t \frac{\sum_{\lambda} \lambda^2 e^{- t \lambda^2}}{\sum_{\lambda} e^{- t \lambda^2}} \;.
\end{align}
This expression runs into a problem for geometries with a non-zero lowest energy mode, since this will dominate the expression at large $t$ and thus hide the large scale structure. 

\begin{figure}
  \begin{subfigure}{0.49\textwidth}
    {\includegraphics[width=\textwidth]{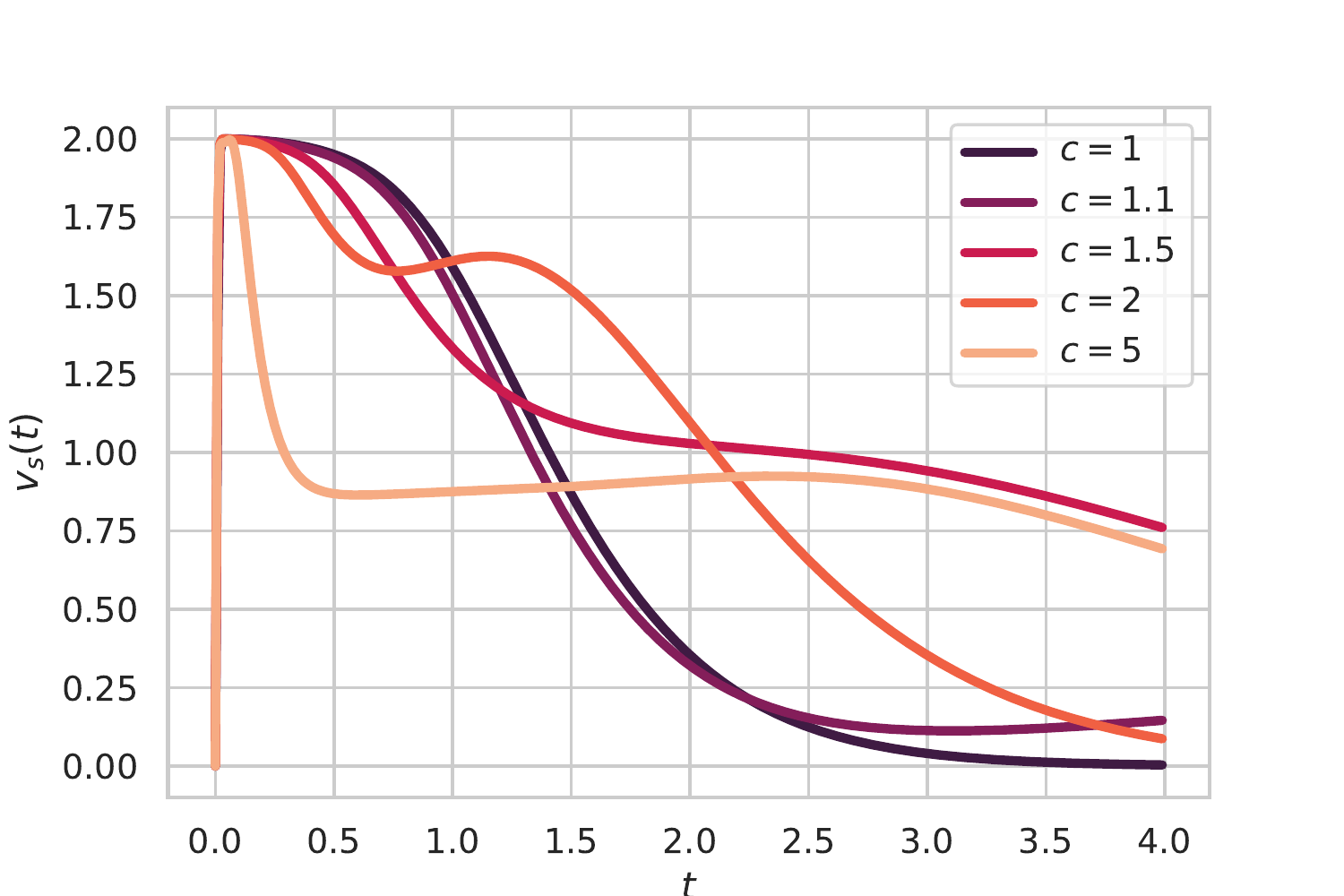}}
    \caption{$c\geq1$}
    \end{subfigure}\hfill
    \begin{subfigure}{0.49\textwidth}
    {\includegraphics[width=\textwidth]{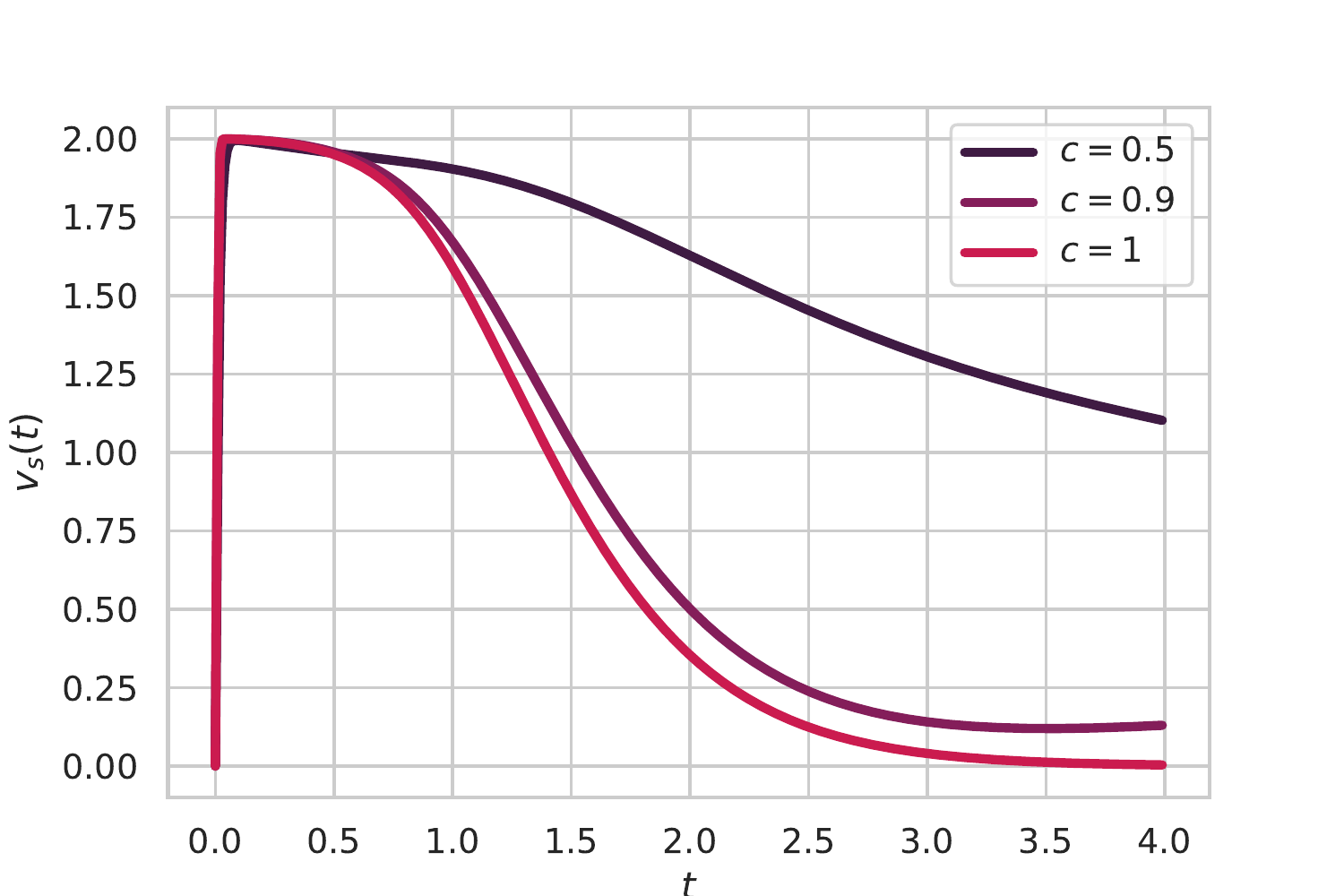}}
    \caption{$c\leq1$}
    \end{subfigure}

    \caption{Spectral variance curve of the different deformed spheres.}\label{fig:spec_var}
\end{figure}

The Dirac operators for fuzzy spaces often have a non-zero smallest eigenvalue, so in~\cite{Barrett_Druce_Glaser_2019} we introduced the spectral variance.
Our studies showed that this new observable works similar to the spectral dimension, but can also usefully be studied for operators that do not have a zero mode
\begin{align}
v_s(t)&= 2t^2 \left(\langle \lambda^4 \rangle -\langle \lambda^2 \rangle^2 \right)  \;.
\end{align}
Calculating this using the spectra for $n=20$ from above we find Figure~\ref{fig:spec_var}.
There we see that, for the deformations with $c>1$ the geometry seems to be approximately $2$ dimensional on short scales, but then appears $1$ dimensional on large scales.
This is consistent with a sphere being narrowed down to a cigar shape, as our scaling analysis above predicts.
The behaviour for $c<1$ is also clear, as $c$ becomes smaller the region that can be approximated as $2$ dimensional becomes larger.
This is explained by considering that $c<1$ corresponds to scaling two axes of the ellipsoid to become longer, thus creating an oblate spheroid.

\subsection{Volume}

\begin{figure}
  \centering
\includegraphics[width=0.8\textwidth]{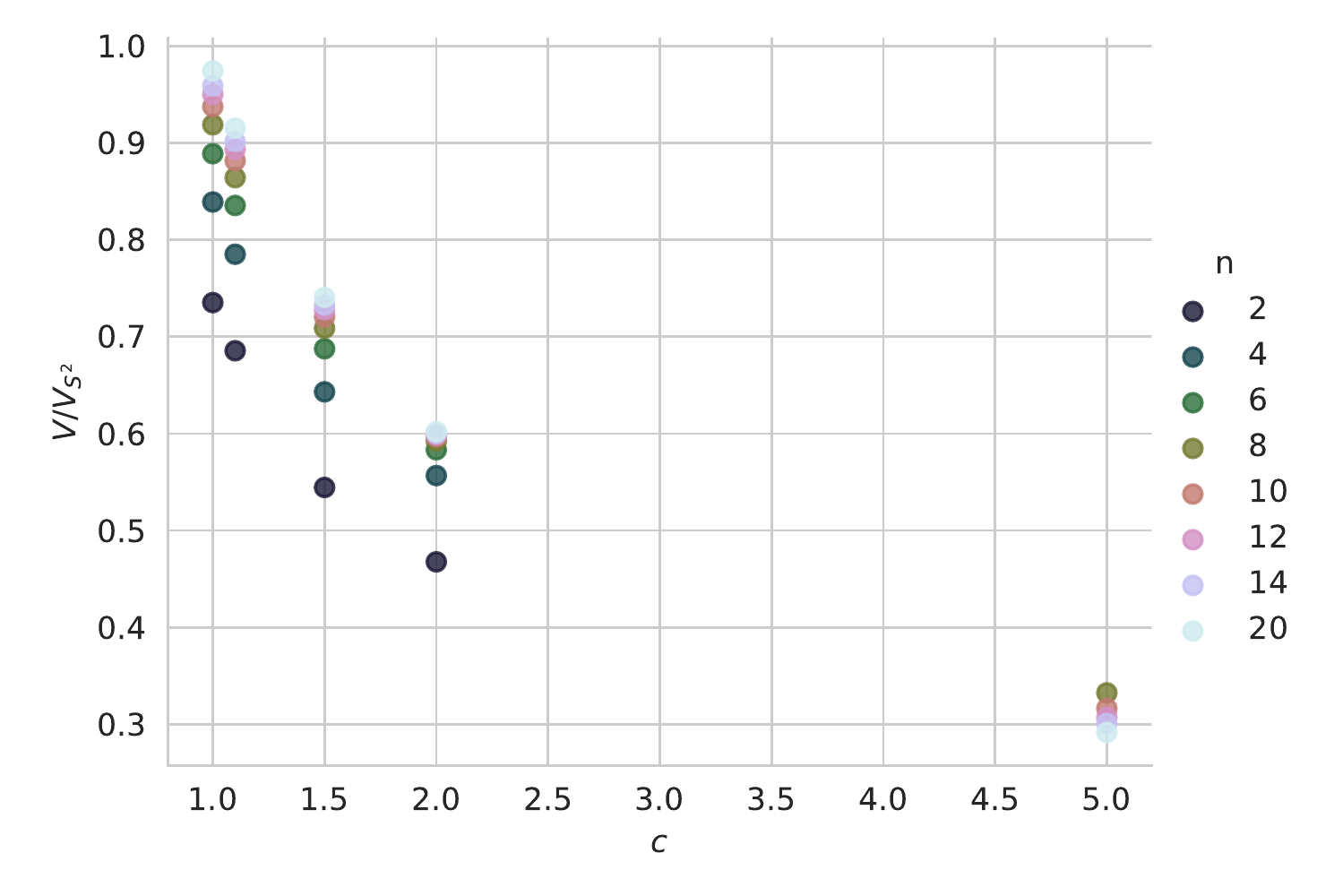}
\caption{\label{fig:vol} This plot shows the change in the volume of the geometry against the deformation parameter $c$, for different matrix sizes $n$.}
\end{figure}

We estimate the volume based on the spectral zeta function as described in~\cite{Stern_2018}, of the space.
\begin{align}\label{eq:vol}
  V_{geom} &= \frac{{(4 \pi)}^{\frac{d}{2}}}{m} {\left( \frac{\left(\log{(\Lambda+1)}\right)^2}{\Lambda} \right)}^{\frac{d}{2}} \sum_{\lambda'} \frac{e^{-\lambda'-1}}{\lambda'+1} \Gamma(1-\frac{d}{2},1) \;,
\end{align}
using rescaled eigenvalues $\lambda'= \frac{\left(\log{(\Lambda+1)}\right)^2}{\Lambda} \lambda$.
How this volume changes as a function of the deformation parameter $c$ for $n=20$ is shown in Figure~\ref{fig:vol}, where we have plotted the volume, divided by the volume of the two sphere\footnote{To be precise we have divided by twice the volume of the two sphere, since the eigenvalue doubling leads to the volume of two spheres appearing here.}.
We see that there is a slight $n$ dependence, but that the value for the round sphere becomes closer to $1$ as $n$ increases, marking this $n$ dependence as a cut-off effect.
This is an effect of the way the fuzzy sphere is normalized, whereby an increase in $n$ leads to an improved resolution of the geometry, instead of a larger surface.
In principle one could change the normalization and generate spheres that keep the resolution fixed, and instead lead to ever larger spheres being described.

\section{Implementing visualization for fuzzy spaces}
In our past work~\cite{Glaser_Stern_2021}, we implemented code to visualize the truncated spectral triple for a continuum sphere.
This code worked by generating localized states, that is states with a small dispersion (defined below in~\eqref{eq:disp}). 
We were able to show that, for truncated spectral triples, these states converge to points in the infinite matrix size limit.
Calculating the Connes distance between these states leads to a distance matrix which was then used to calculate an embedding for the points using a weighted SMACOF algorithm\footnote{SMACOF stands for ``Scaling by MAjorizing a COmplicated Function''.} for stress majorization.

In the following we will explain how this code works, and at the same time elaborate those points of the code that need to be modified to generalize this code to fuzzy spaces in general, and the deformed fuzzy sphere in particular. This covers three points in particular\footnote{For the most general fuzzy spaces we would also need to consider how to choose the dimension of the space the geometry is embedded in, however for the deformed fuzzy sphere this will be fixed at $d=3$, and so we do not discuss it here.}:
\begin{itemize}
  \item We require proxies for the coordinates to be able to calculate the dispersion, and generate localized states that are roughly evenly distributed on our geometry.
  \item We need to determine the number of points we want to generate, this will depend on the cut-off of the geometry, the volume of the geometry and the dimension of the fuzzy space.
  \item After generating the states we will calculate the Connes distance between them. To do this we require a basis for the algebra to optimize over.
\end{itemize}
We will cover these  problems in the following subsections.
The code for the deformed fuzzy sphere visualization, where we implemented these changes, is available under~\cite{github_code}.

\subsection{The proxies for the coordinate calculation}
The first step in visualizing a spectral triple is to generate states of low dispersion, which we showed can, in the context of truncated spectral triples, be understood as well localized.
The dispersion for a state $s$ is defined as 
\begin{align}\label{eq:disp}
 \delta(s) = \sup_{a\in \mc{A}} \left\{ |s(a^2) - {s(a)}^2| \big | |[D,a]| \leq 1 \right\} \;.
\end{align}
While this is a good definition for a given state, using it in code that tries to find states of minimal dispersion leads to a convex double optimization; we optimize $s$ to minimize $\delta(s)$, but at the same time each $\delta(s)$ optimizes $a$ to maximize it.
This type of problem is notoriously hard to solve numerically.
It is thus necessary to use a proxy, we used the projector $P$, which arises from the Heisenberg relation defined in~\cite{Chamseddine_Connes_Mukhanov_2014}.
This gives rise to coordinate matrices $\{X,Y,Z\}$ that give us an approximation of the dispersion
\begin{align}\label{eq:disp_proxy}
  \tilde{\delta}(s) =  \left\{ |\sum_{W \in \{X,Y,Z\}} s(W^2) - {s(W)}^2| \right\} \;.
\end{align}
We were able to show that, while $\delta$ and $\tilde \delta$ do not need to be numerically close, they will show the same behaviour, and in particular minimize on similar states, at least for the case of truncated geometries and the limit of large truncations.

Minimizing the dispersion leads to a state that is well localized, however generating an ensemble of points that only minimize the dispersion will lead to points that cluster together, due to the computer algorithm.
To cover the entire sphere we added a repulsive potential between the states.
We then minimize the dispersion proxy $\tilde \delta$ plus a Coulomb potential
\begin{align}
E(s) = \tilde \delta(s) + \sum_{i} \frac{g}{\tilde{d} (s_i,s)}
\end{align}
where the sum goes over all states already generated, and the distance proxy $\tilde{d}$ is 
\begin{align}\label{eq:dist_proxy}
  \tilde{d}(s_i,s_j) =  \left\{ |\sum_{W \in \{X,Y,Z\}} s_i(W) - s_j(W)| \right\}.
\end{align}
We use the distance proxy instead of the Connes distance defined in~\eqref{eq:ConnesDist}, since this would also lead to a double optimization problem.
Again we were able to show that this is a good proxy for the distance, for the truncated spectral triples.
We can then generate a set of points by successively minimizing $E(s)$ a set number of times.

For fuzzy spaces the matrices $\{X,Y,Z\}$ can be approximated through the matrices $L_i$ arising in the Dirac operator.
This is supported by other work on visualizing fuzzy spaces (without the entire machinery of the spectral triple attached) in~\cite{Schneiderbauer_Steinacker_2016}.
It is also the simplest solution, since these matrices are the only data available to us, without adding additional input into our system.
For the present article we thus effectively use
\begin{align}\label{eq:fuzzyProxy}
  \tilde{\delta}(s) =  \left\{ |\sum_i s(L_i^2) - {s(L_i)}^2|  \right\} \\
 \tilde{d}(s_1,s) =  \left\{ |\sum_i s_1(L_i) - s(L_i)|  \right\} \;.
\end{align} 

\subsection{The number of points and dimension}
The number of points that can be generated depends on the resolution of the truncated space or the matrix size of the finite spectral triple.
Our code determines the number of localized states to search, based on the dispersion of the states obtained $\delta$, and the overall volume of the space $V_{geom}$.
The maximal number of states we generate is given as $V_{geom}/( \delta^{d}B_d)$, where $B_d$ is the volume of unit sphere in $d$ dimension. 
To calculate this we need to estimate the dimension, for which we use the spectral dimension as described in~\cite{Barrett_Druce_Glaser_2019}, and the volume as already discussed above in~\eqref{eq:vol}

To estimate the dimension of the geometry we evaluate the spectral dimension at the point $t_d=\frac{\left(\log{(\Lambda+1)}\right)^2}{\Lambda}$, where $\Lambda$ is the largest absolute value among the eigenvalues.
This value is chosen based on~\cite{Stern_2018} to lie above the region where the finite size of the matrices starts to influence the result, but well below the region where the large scale structure of the geometry will start to interfere.
Since this does not necessarily lead to an integer value for the dimension, we round the result to the closest integer.

\begin{figure}
  \begin{subfigure}{0.46\textwidth}
    {
    \includegraphics[width=\textwidth]{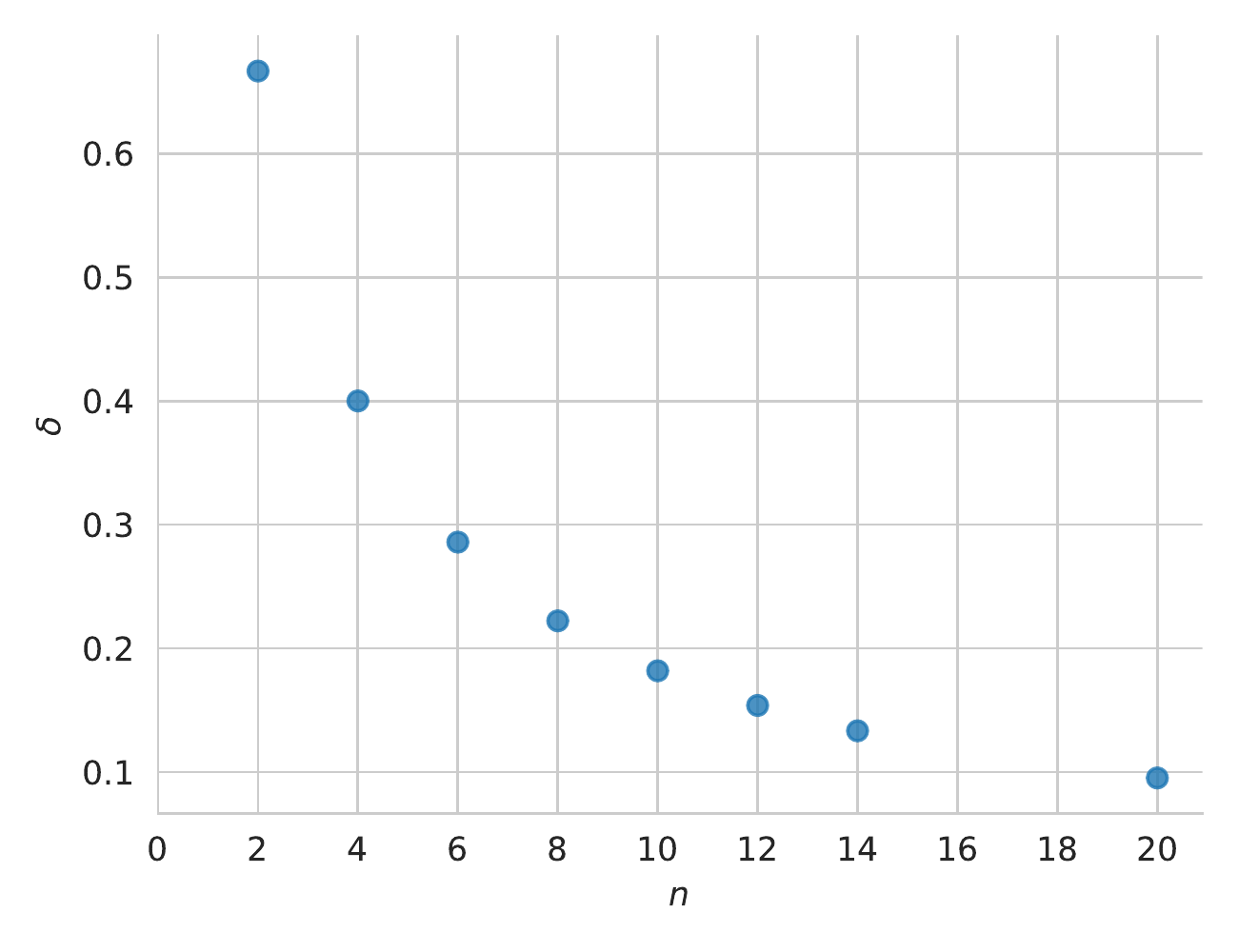}}
    \caption{Dispersion\label{fig:disp}}
  \end{subfigure}
  \begin{subfigure}{0.53\textwidth}
  {
  \includegraphics[width=\textwidth]{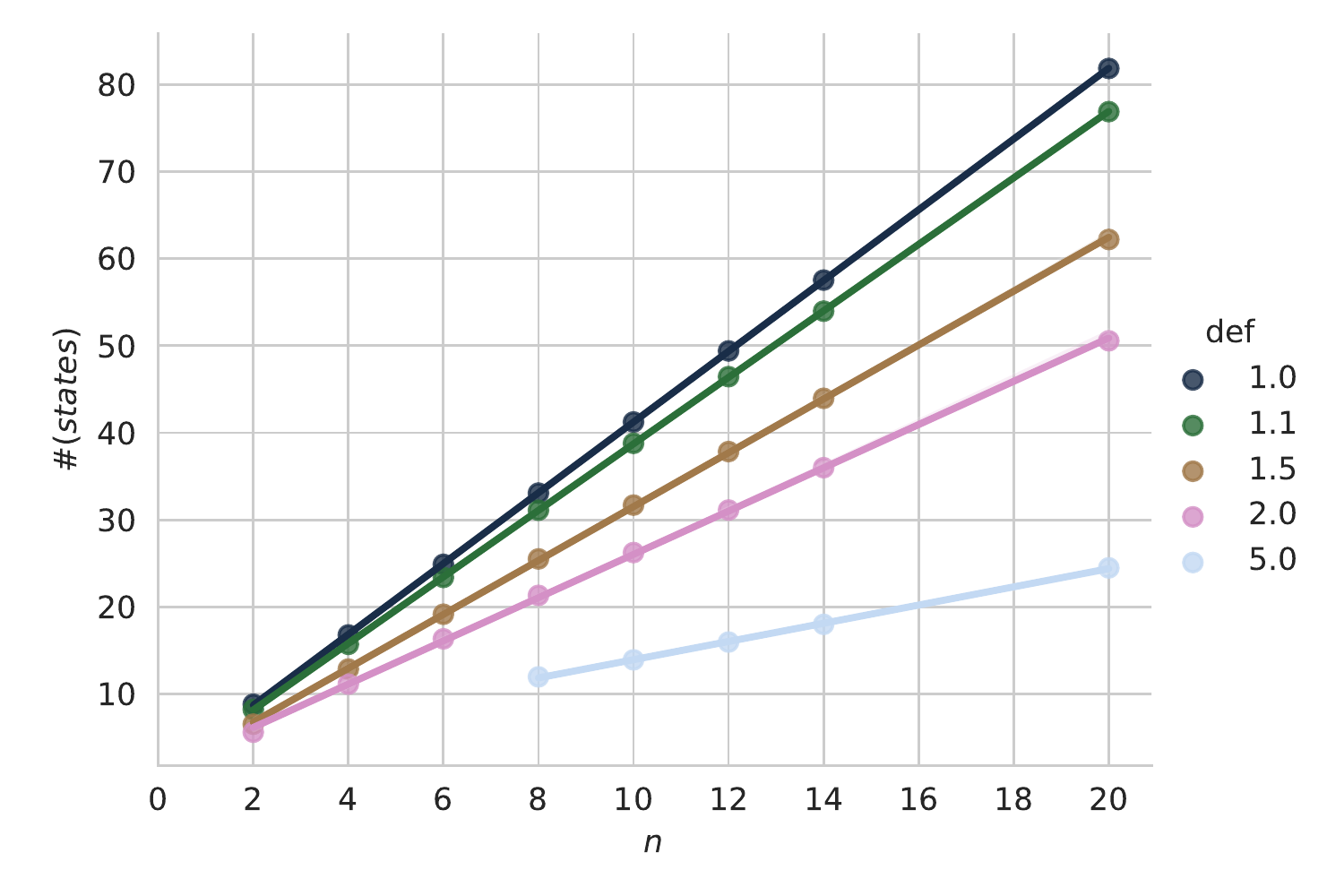}}
  \caption{Maximal state number\label{fig:states}}
\end{subfigure}

\caption{
  The left-hand plot shows the change in the dispersion with $n$.
The right-hand image shows the estimated number of states as a function of $n$.
The colour indicates the deformation parameter used, clearly each deformation has a different linear dependence.}
\end{figure}

The other ingredient to estimate the number of states is the dispersion $\delta$ of the states.
In~\cite{Glaser_Stern_2021} this was estimated based on the eigenvalue cut-off of the geometry as $ \log{(\Lambda_{cutoff})}/\Lambda_{cutoff}^2$.
While this works well for truncated continuum geometries it does not generalize to fuzzy spaces.
A more universal (and precise) way to approximate the maximal number of non-overlapping states for the geometry is to use the dispersion $\delta$ of the generated states from the code.
The behaviour of this dispersion as a function of the matrix size $n$ is illustrated in Figure~\ref{fig:disp}.
The dispersion of the states is independent of the deformation, since the algebra of generators is the same for all deformation parameters.

So we find that the dispersion of the states depends on $n$, but not on the deformation of the geometry, while the volume, as we saw above, is almost independent of $n$, but strongly changes with the deformation.
The dimension of the geometry is stable at $2$, except for geometries with very low matrix size, where the resolution of the geometry is not sufficient to show the $2$d dimensional nature of the deformed sphere.
We exclude these in our plot of the number of expected states, which we show in  Figure~\ref{fig:states}.
Comparing it to Figure~\ref{fig:vol} above we can see that the change of the volume with $c$ is what drives the change in the maximal number of states.

This image clearly shows that the number of states rises linearly with $n$, which, from our equation to calculate the maximal number of states, implies that the dispersion of the states falls of like $n^{-2}$.
Unfortunately the computational complexity per state rises fast.
A test run showed that the generation of two states and evaluation of their distance scales strongly with $n{}$\footnote{A short exploratory run was not able to determine whether it scales like a higher power law, or exponentially.}, and the number of pairs whose distance needs to be evaluated scales like $n^2$.
So even though the number of states does not rise quickly, the time to calculate them does, thus limiting us in the size of matrices we can use.
However, as we will show below even for $n=8$ the algebra contains enough information to lead to impressive visualizations of the geometries. 

\subsection{Choosing a basis for the algebra}
After having generated the states, we need to calculate their distance, using the Connes distance function.
Since the states are fixed at this point we can optimize the exact expression, to do this we use a basis of the algebra $\mc{A}$.
For the truncated sphere the natural basis of the algebra is formed by a truncated matrix representation of the spherical harmonics, which we used in our past work~\cite{Glaser_Stern_2021} to calculate the Connes distance between states.

In the case of the deformed sphere we are not given a basis a priori, however we are handed the generators of the algebra, in the form of the algebra elements $L_i$ that are part of the Dirac operator.

We then generate a basis for the space of all algebra elements by calculating all products of the $L_{i}$, applying the Poincaré-Birkhoff-Witt theorem.
Since the $L_{i}$ for the deformed sphere satisfy the usual commutation relations we can use this to reduce the number of states to calculate.
We also check each generated element and do not keep it if it is linearly dependent on any already generated element.

For more general finite spectral triples this will not generally be true.
However, it is possible to choose the most naive basis on $\Mn$, using all matrices with only one non-zero entry, and find comparable results.
This choice of algebra is also implemented as an option in our code.

\section{Images of a deformed fuzzy sphere}
\begin{figure}
  \begin{subfigure}{0.49\textwidth}
    {\includegraphics[width=\textwidth]{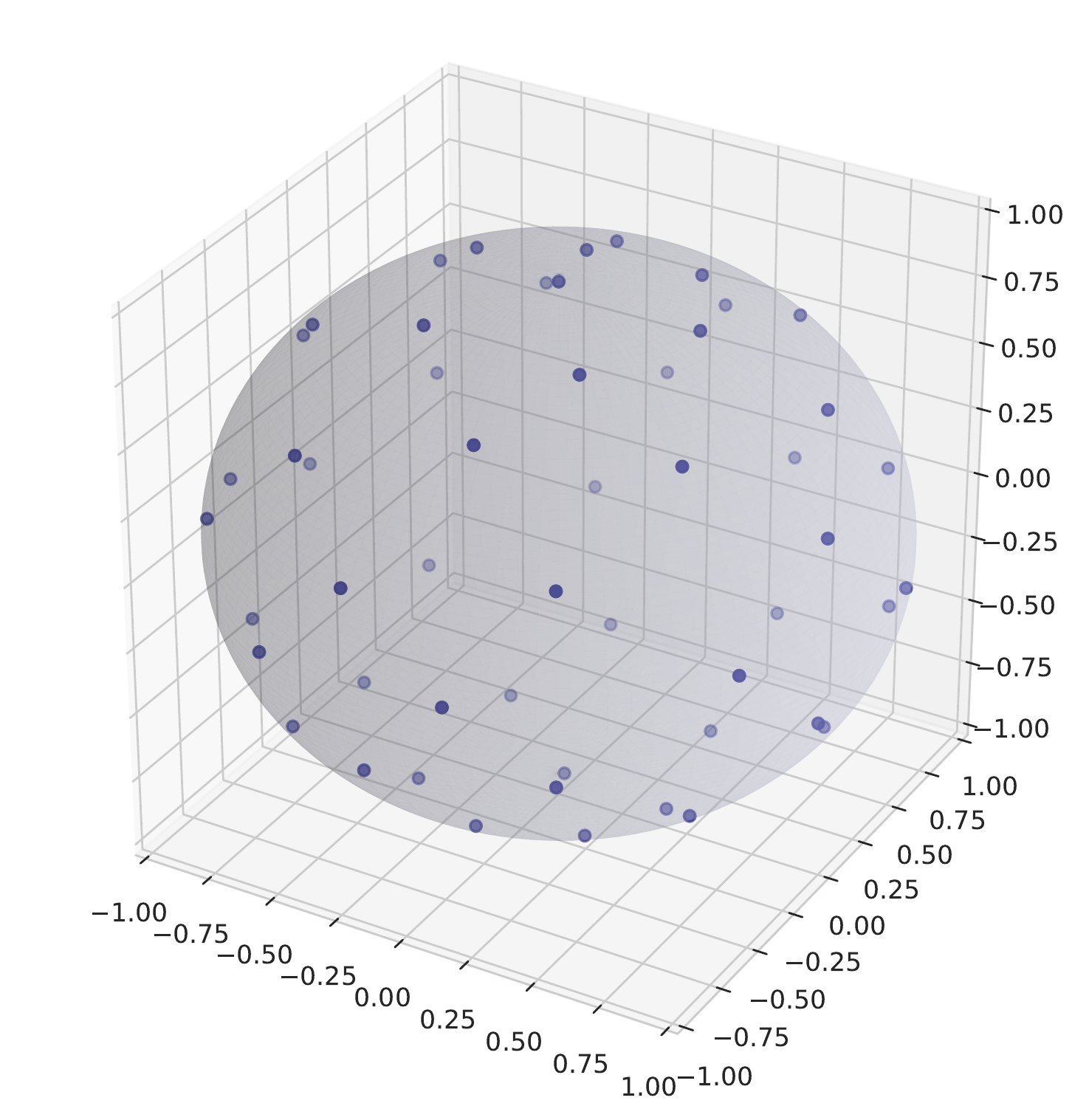}}
    \caption{$c=0.5$}
    \end{subfigure}\hfill
    \begin{subfigure}{0.49\textwidth}
    {\includegraphics[width=\textwidth]{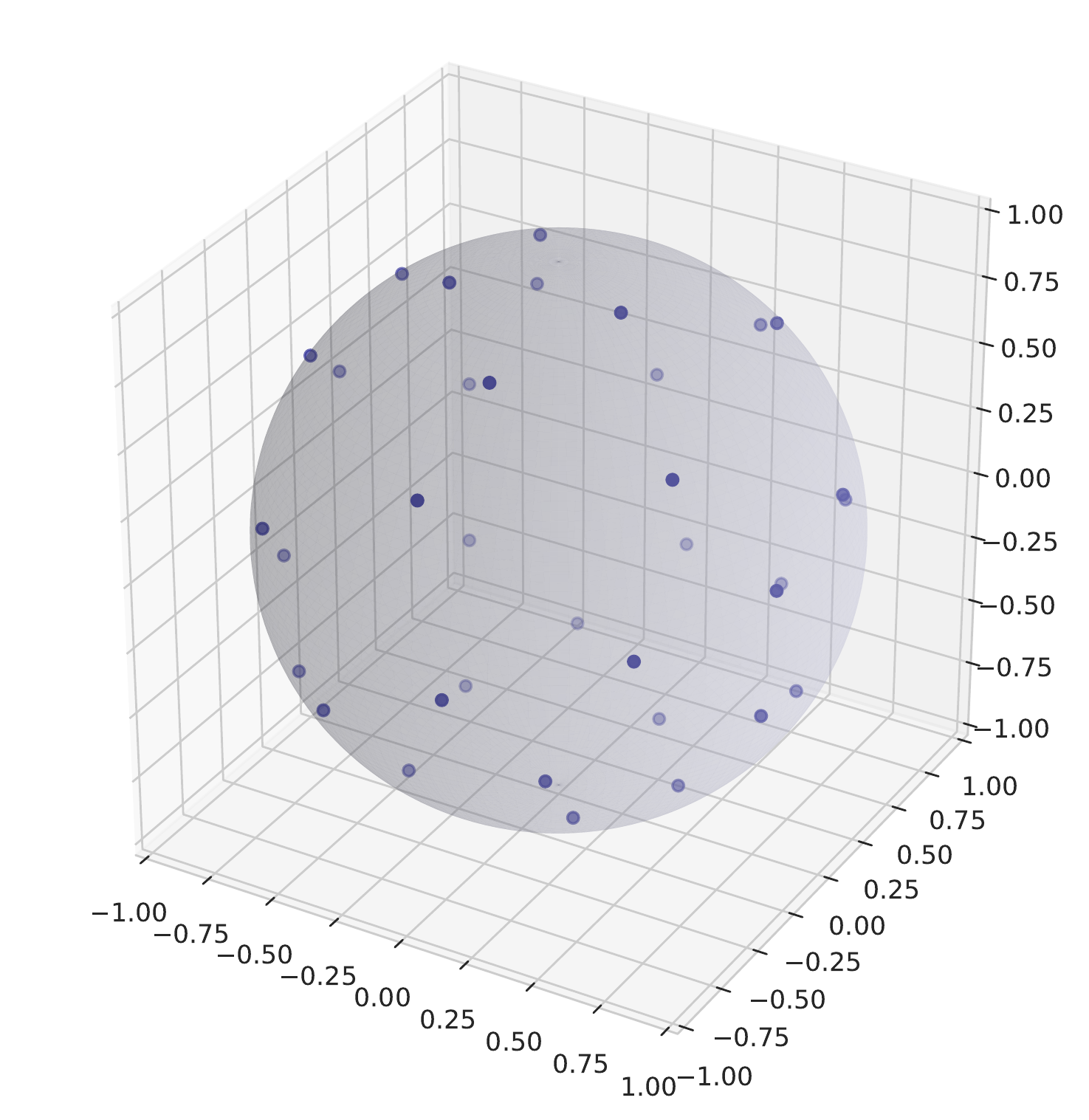}}
    \caption{$c=0.9$}
\end{subfigure}
  \begin{subfigure}{0.49\textwidth}
{\includegraphics[width=\textwidth]{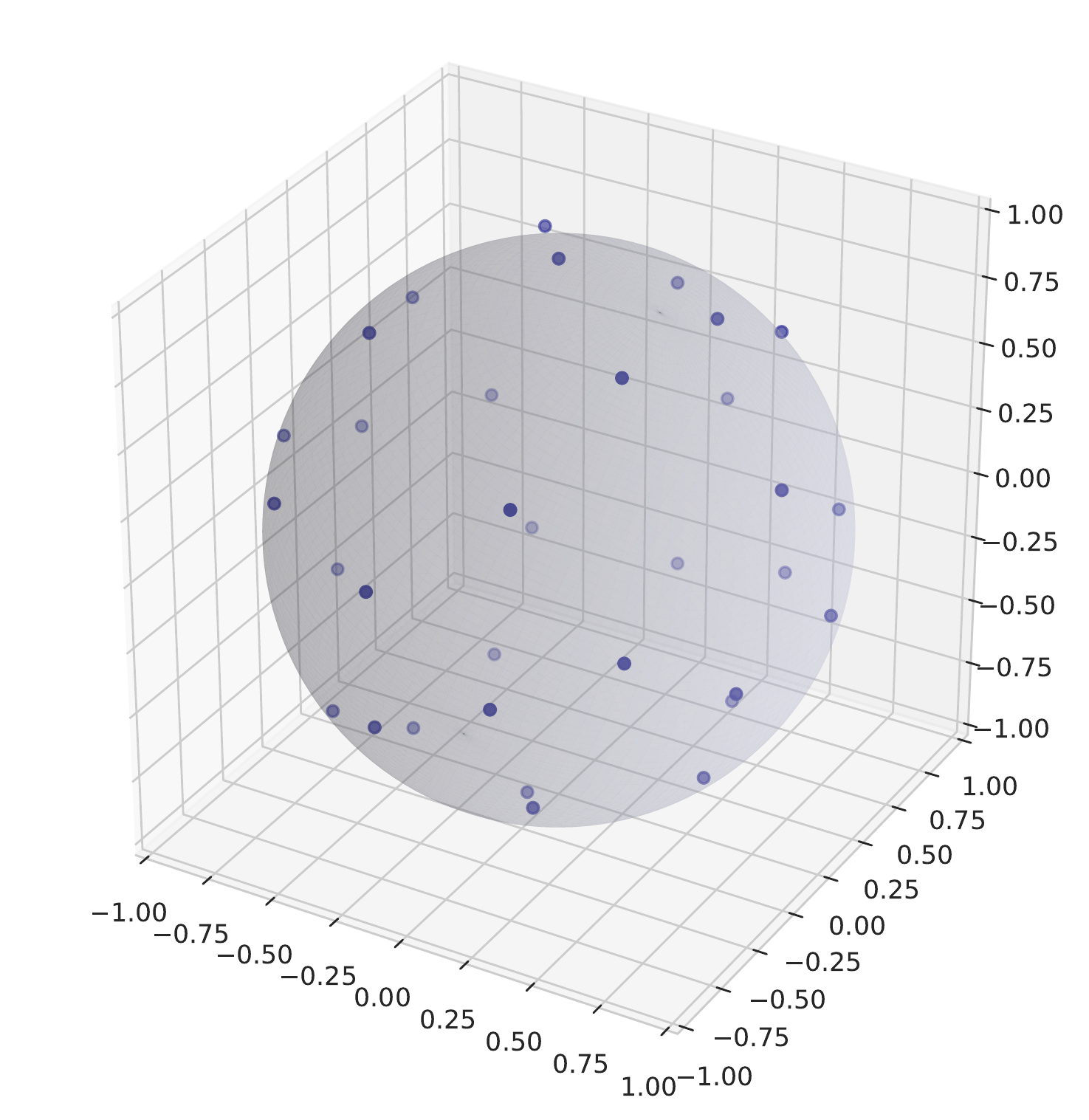}}
\caption{$c=1.0$}
\end{subfigure}\hfill
\begin{subfigure}{0.49\textwidth}
{\includegraphics[width=\textwidth]{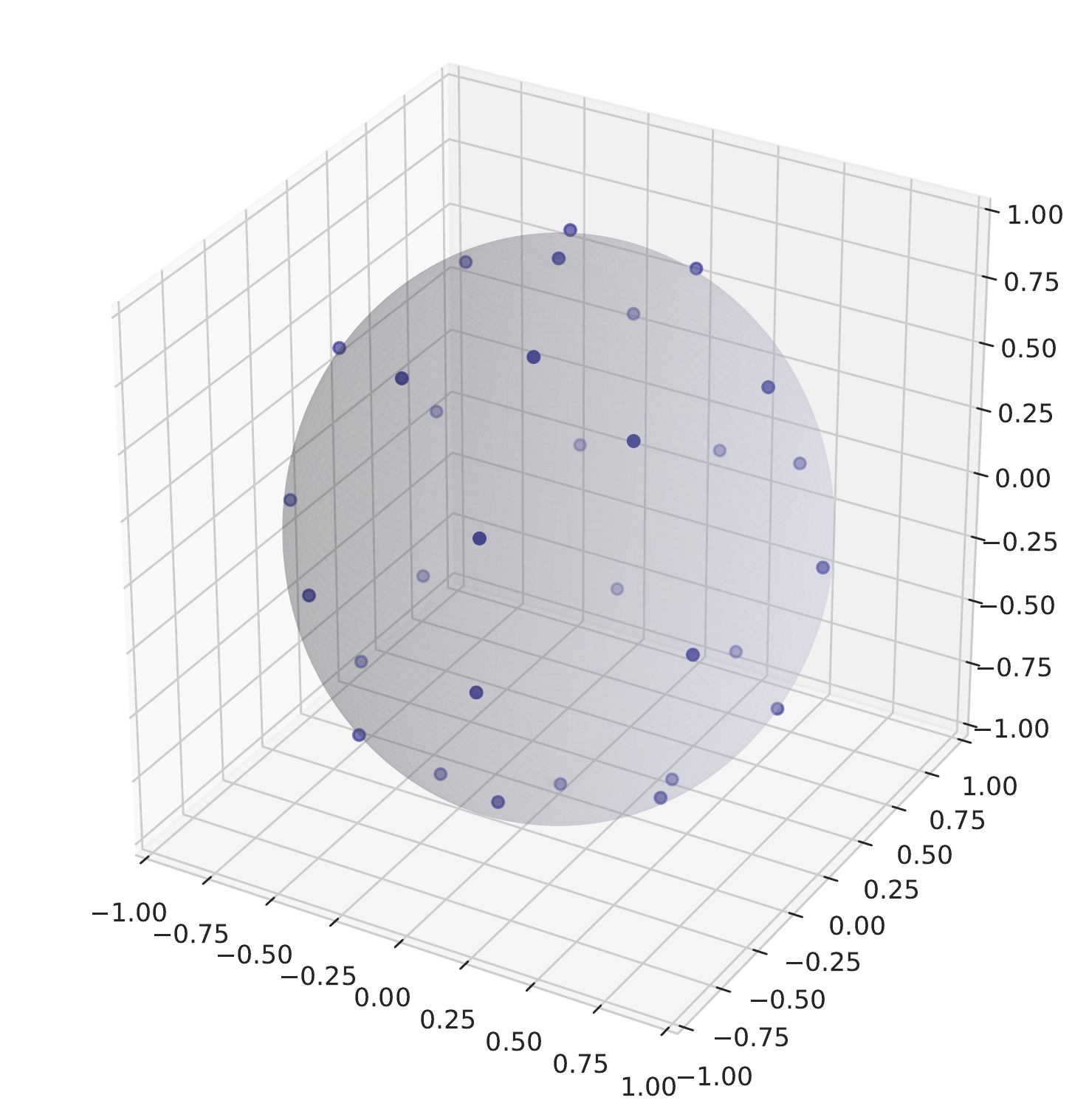}}
\caption{$c=1.1$}
\end{subfigure}

\caption{Embeddings of the deformed spheres with their best fit ellipsoids plotted for comparison.}\label{fig:elfit}
\end{figure}

\begin{figure}
\begin{subfigure}{0.49\textwidth}
  {\includegraphics[width=\textwidth]{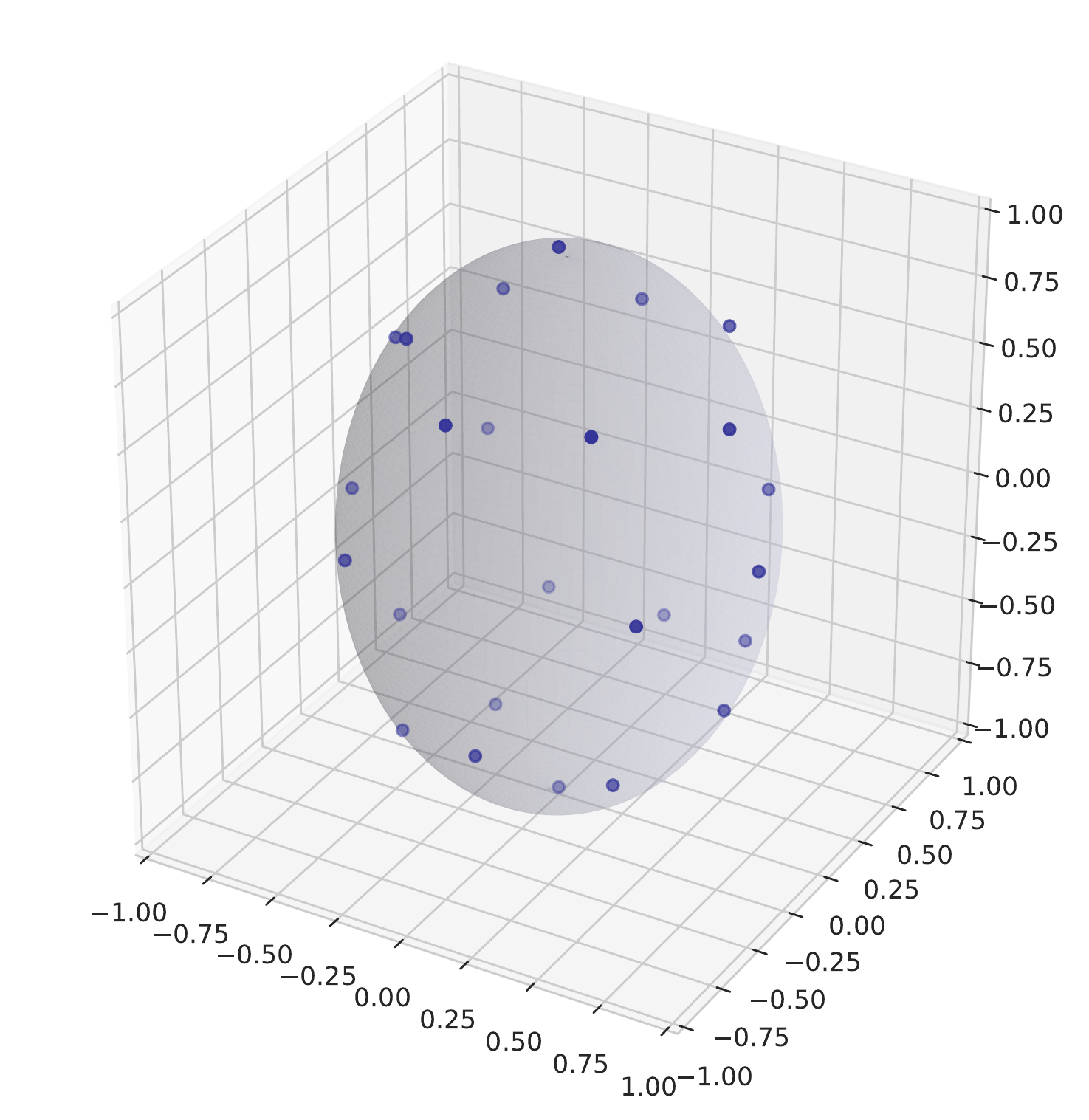}}
  \caption{$c=1.5$}
  \end{subfigure}\hfill
  \begin{subfigure}{0.49\textwidth}
  {\includegraphics[width=\textwidth]{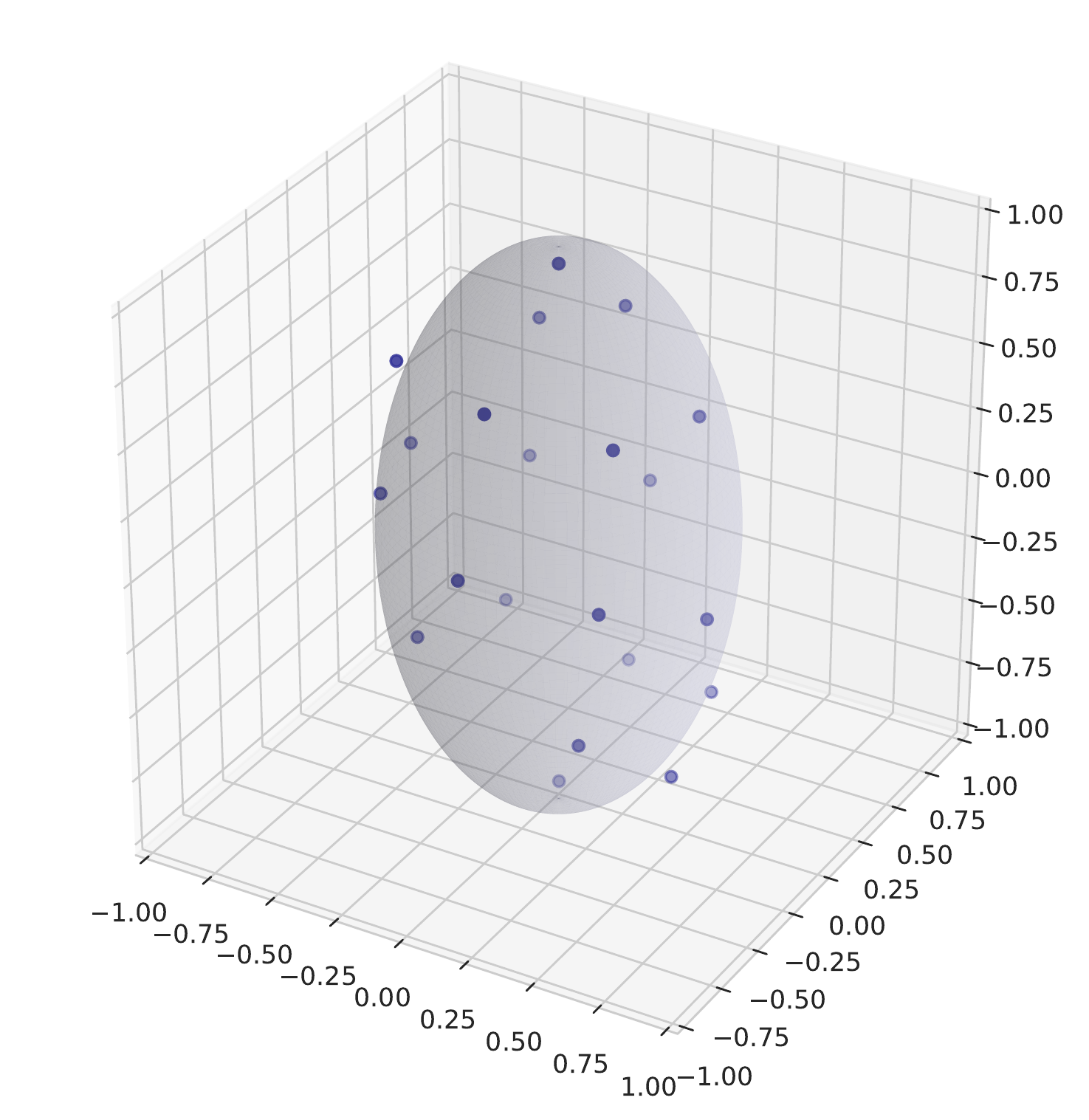}}
  \caption{$c=2.0$}
\end{subfigure}

\centering

\begin{subfigure}{0.645\textwidth}
\includegraphics[width=\textwidth]{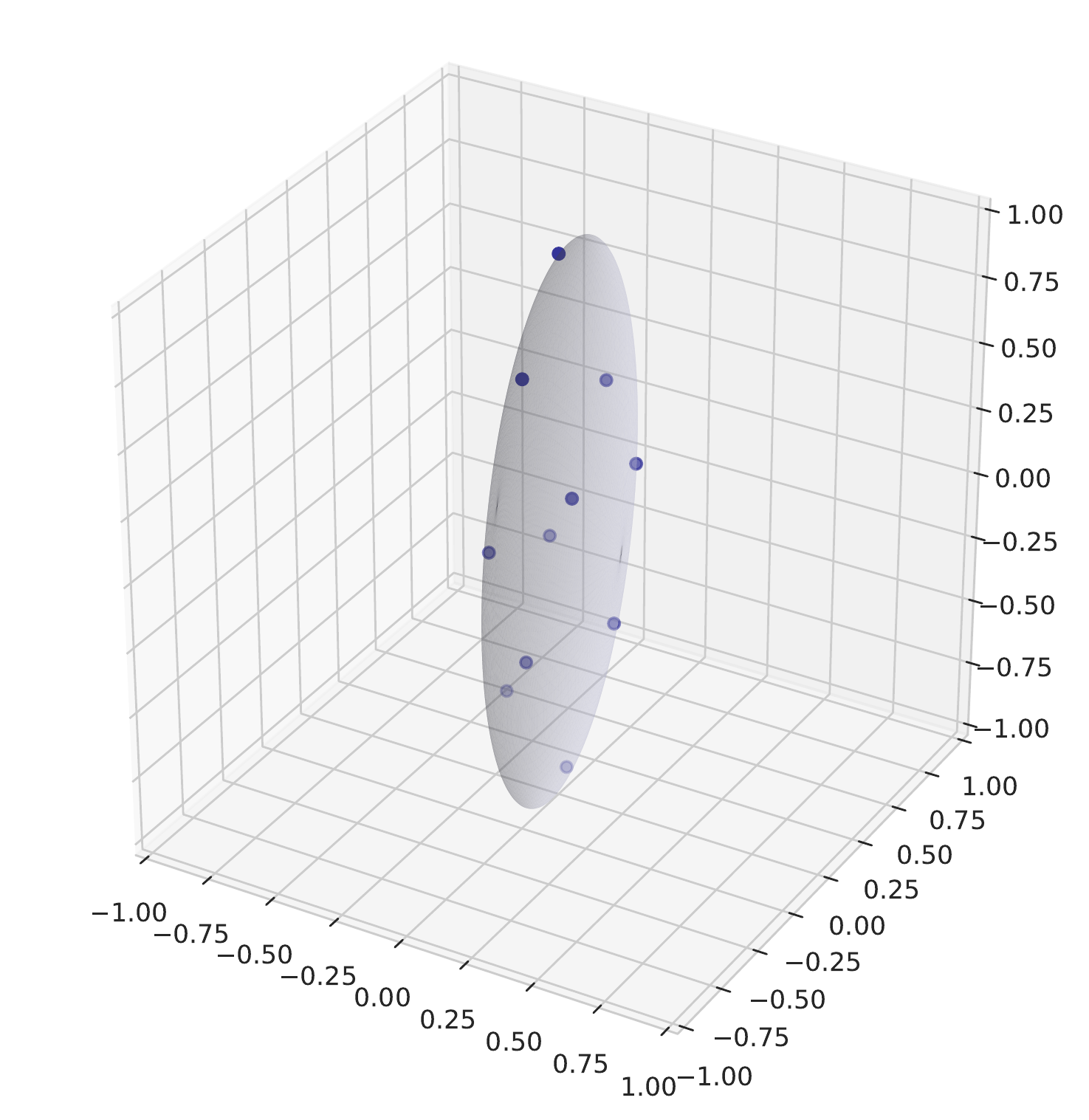}
  \caption{Best fit ellipsoid for the strongest deformation $c=5$}\label{fig:elfitd5}
\end{subfigure}
\caption{Embeddings of the deformed spheres with their best fit ellipsoids plotted for comparison.}\label{fig:elfit2}
\end{figure}

To test the code we generated embeddings of deformed fuzzy spheres.
We will first discuss the constrained case with $c_0,c_{12},c_{13}$ at $1$ and $c_{23}=c$ at the end of the section observe some more general deformations.
The results would be equivalent if we had changed one of the other two directions and kept this one constant.
The embedding for these deformed fuzzy spheres for $n=8$ are shown in Figures~\ref{fig:elfit} and~\ref{fig:elfit2}.

To understand these images better we have fit ellipsoids to the embedding data.
These fits have five degrees of freedom, two angles of rotation for the main axis\footnote{Since the embedding algorithm starts with a random point, the direction of the embedding is random.}, and the three deformation parameters $a_k$.
Since we fit the three deformation parameters, the radius can be held fixed at $1$, as a change in radius is simply a simultaneous rescaling of all axes by the same amount.
For the fit we optimized a function $f$ which was the sum of the square of the differences from each point to the fitted ellipsoid, and the fit itself done via pythons optimize library~\cite{2020SciPy-NMeth}.

In Table~\ref{tab:simple fits} we compare the results for these fits with our expected deformations from equation~\eqref{eq:defo_expl}. 
We find that they do agree reasonably well, taking into account that, especially for $c_{13}=5$ we are fitting $5$ free parameters to $11$ points.
The fits for the rounder states, where more points are available are thus better.

The data works equally well for $c<1$ as for $c>1$ which was to be expected.
The difference between $c<1$ and $c>1$ is that the former leads to oblate spheroids, while the latter leads to prolate spheroids.

\begin{table}
  \caption{Best fit values for the main axes of the ellipses.}\label{tab:simple fits} 
  \centering
  \begin{tabular}{r r  r r  r}
    \toprule
  $c$ &  expected axes & best fit axes & angle of axis & $f$ / d.o.f. \\  \midrule 
  $ 0.50$ & $ (1.00,2.00,2.00)$ &$ (0.96,1.18,1.19)$ & $ (0.00,3.14)$ &$ 0.0020  $  \\ 
  $ 0.90$ & $ (1.00,1.11,1.11)$ &$ (1.00,1.01,1.03)$ & $ (-0.00,0.83)$ &$ 0.0024  $  \\ 
  $ 1.00$ & $ (1.00,1.00,1.00)$ &$ (0.98,0.99,0.99)$ & $ (0.39,0.00)$ &$ 0.0015  $  \\ 
  $ 1.10$ & $ (0.91,0.91,1.00)$ &$ (0.92,0.93,1.01)$ & $ (0.01,0.39)$ &$ 0.0008  $  \\ 
  $ 1.50$ & $ (0.67,0.67,1.00)$ &$ (0.73,0.75,1.03)$ & $ (0.03,0.21)$ &$ 0.0025  $  \\ 
  $ 2.00$ & $ (0.50,0.50,1.00)$ &$ (0.58,0.61,1.06)$ & $ (-0.00,0.59)$ &$ 0.0224  $  \\ 
  $ 5.00$ & $ (0.20,0.20,1.00)$ &$ (0.24,0.25,1.13)$ & $ (1.67,-0.00)$ &$ 0.0358  $  \\ 
  \bottomrule
\end{tabular}
\end{table}

\subsection{Correlation coefficient}
Another tool to check our visualization, is to compare how well the distance between the states in the embedding agrees with the distance of the points calculated according to the spectral distance.
The easiest way to do so, is to observe the correlation coefficient between these two quantities.
If the constellation of points does not embed into the flat space provided for the embedding the algorithm will try to do its best, but will not be able to embed all points well, and thus the correlation coefficient will be high for some points.
As an example for this we have run an embedding of the fuzzy $S^2$ into $2$d space.

\begin{figure}
  \centering
\includegraphics[height=0.5\textwidth]{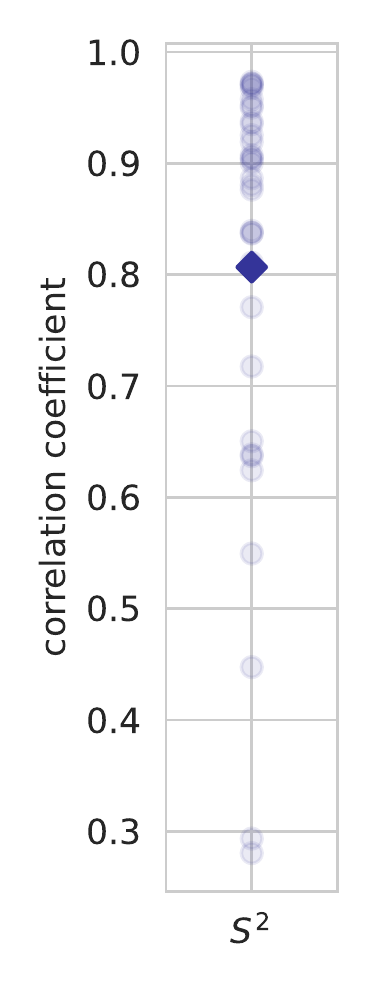} \hspace{30pt}
\includegraphics[height=0.5\textwidth]{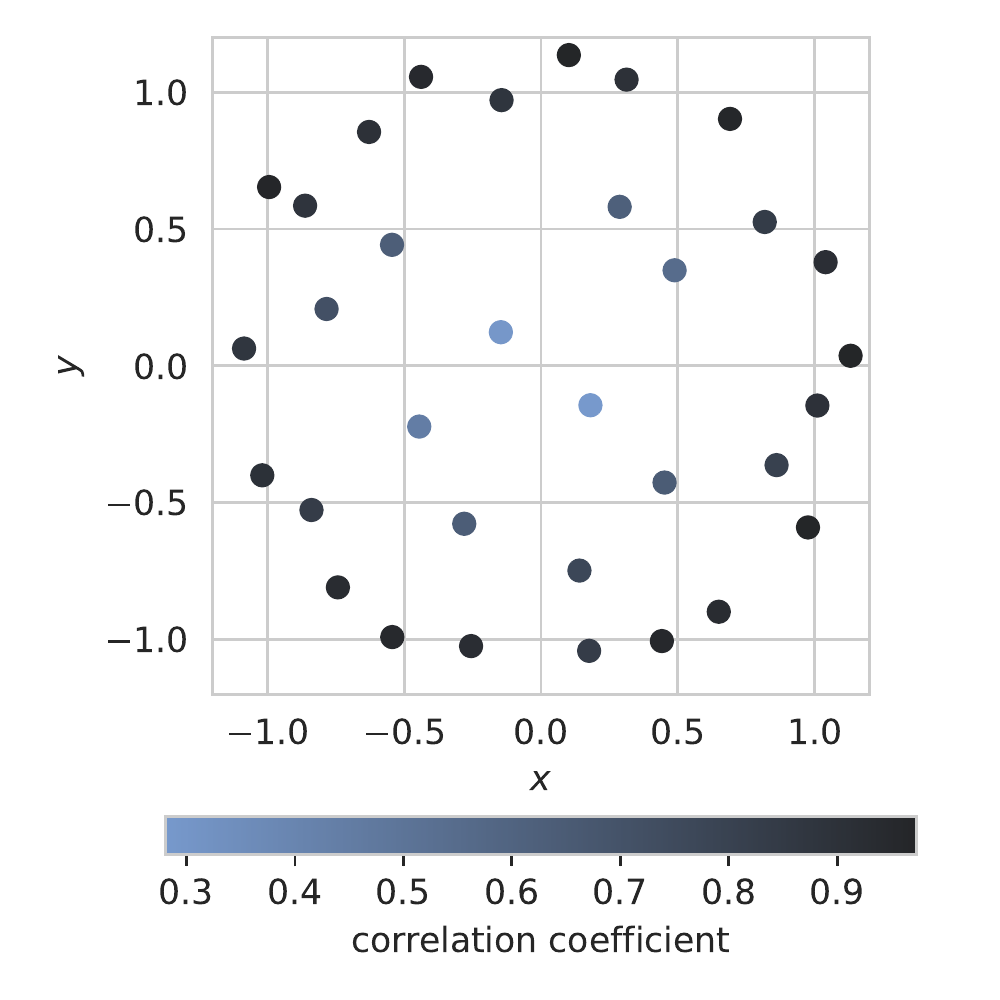} 
  \caption{To demonstrate the correlation coefficient for a failed embedding we have here embedded the fuzzy $S^2$ without deformation into $2$d.
  While the right-hand image of the embedding seems reasonable, the left-hand image of the correlation coefficients shows us that this embedding did not work well.}\label{fig:cc2d}
\end{figure}

In the left-hand image of Figure~\ref{fig:cc2d} we show the scatter of the value of the correlation coefficient, which in this example ranges from $0.3$ to $1$.
It is clear that while the algorithm is able to match the distances well for some points, it eventually finds points that can not be reconciled with all the others, leading to low values.
In the right-hand image we plot the $2$d embedding, and colour the points by their correlation coefficient.
It is then clear that the points that lie in the middle of the circle do not embed well.

\begin{figure}
  \centering 
\includegraphics[width=0.8\textwidth]{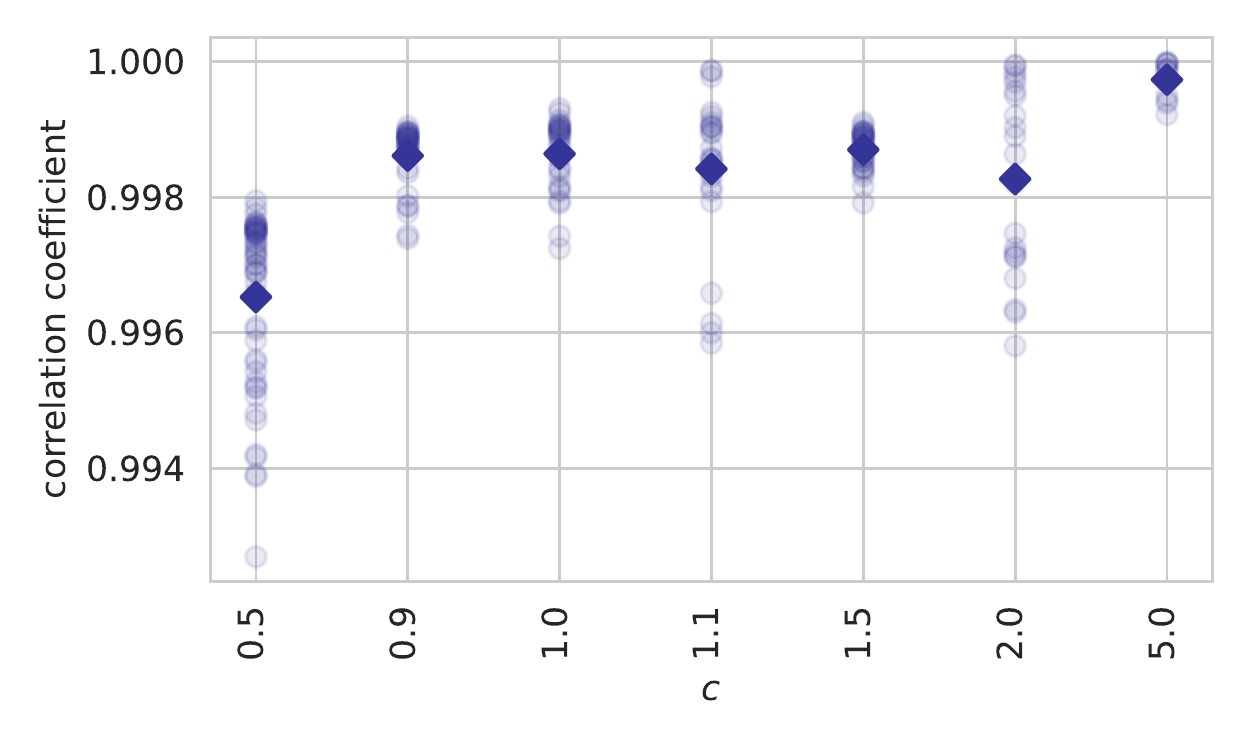}
\caption{Correlation between the distances in the embedding and the distance between the states. The diamond marks the average value, while the transparent points show the values for each individual point.}\label{fig:cc_lineplot}
\end{figure}

In Figure~\ref{fig:cc_lineplot} we show the correlation coefficients for the embedding into $3$d.
Here the diamond shape marks the average value for a given $c$, while the transparent points in the background show the values for each individual point, and thus illustrate that there is little spread.

\subsection{Distance histograms}
The correlation coefficient reduces the data to one convenient number, however it only tells us how well the embedding reproduces the distances we have calculated.
What we would like to do is find some more quantities to, at least qualitatively, compare how well the geometry we find agrees with an ellipsoid.
One such quantity is the collection of the distances between all pairs of points, which can be plotted as a histogram.

\begin{figure}
 \centering 
 \includegraphics[width=\textwidth]{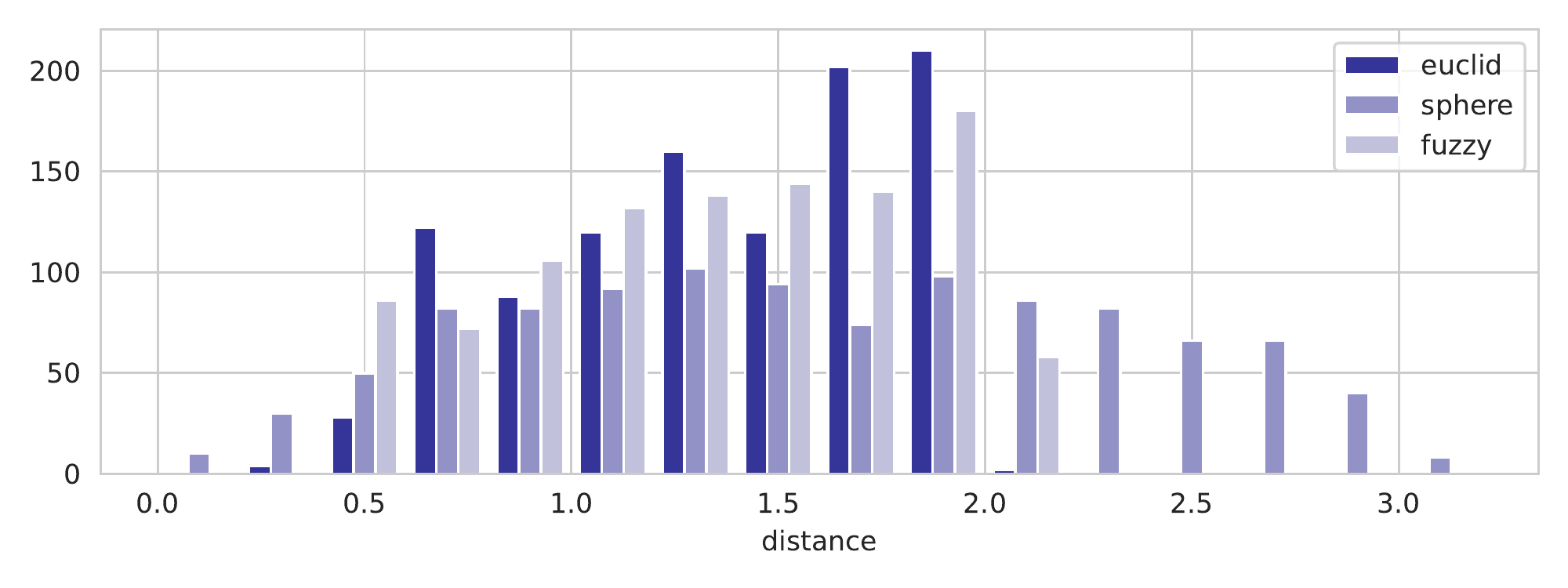}
  \caption{Histogram of the distance between points for the round sphere, comparing the distances found using the Euclidean distance measure, as well as those measured along the sphere with those found from the Connes distance for our embedded fuzzy spaces.}\label{fig:euc_vs_sphere} 
\end{figure}

This is also sensitive to the question whether the Connes distance measure finds the Euclidean distance in the embedding space, or the distance along the sphere.
In Figure~\ref{fig:euc_vs_sphere} we show the distance distribution for the round $c=1$ case, compared to the distance distribution of evenly distributed points for a sphere calculated using the Euclidean distance on $\mathbb{R}^3$ or the distance on the sphere.

This figure shows clearly that the Connes distance reproduces the Euclidean distance between points and not the distance on the sphere.
To ponder why this might be the case we go back to the construction for the Dirac operator in~\cite{Grosse_Prešnajder}.
The construction for the Dirac operator for the fuzzy sphere does use the $3$ dimensional embedding space of the two sphere, which is restricted to the surface of the sphere by fixing the radius. 
It then seems that Connes distance function on the finite spectral triple does not only include points on the surface of the sphere, but also those in the embedding geometry.
This question deserves some further study.

To compare our data for the deformed sphere we can take a sampling of points on an ellipsoid, which was achieved by minimizing a repulsive potential, to generate a random but uniform distribution of points, similar to what we expect to find from our sampling algorithm for the fuzzy space.
As explained above we use an ellipsoid with one axis of length $1$ and two axes rescaled to $1/c$. 

\begin{figure}
  \centering 
  \begin{subfigure}{0.85\textwidth}
    {\includegraphics[width=\textwidth]{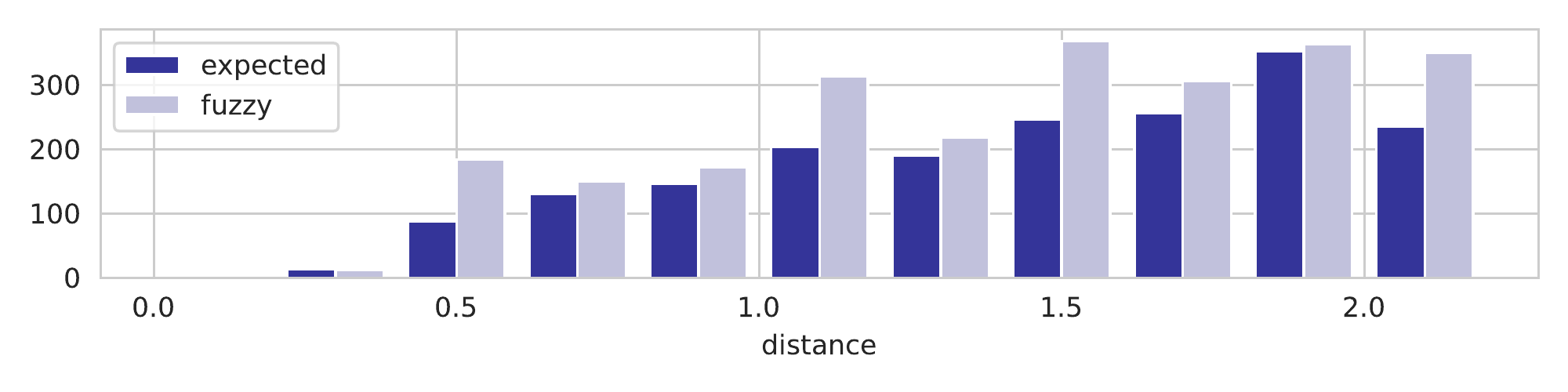}}
    \caption{$c=0.5$ $N=52$}
    \end{subfigure}
    
    \begin{subfigure}{0.85\textwidth}
    {\includegraphics[width=\textwidth]{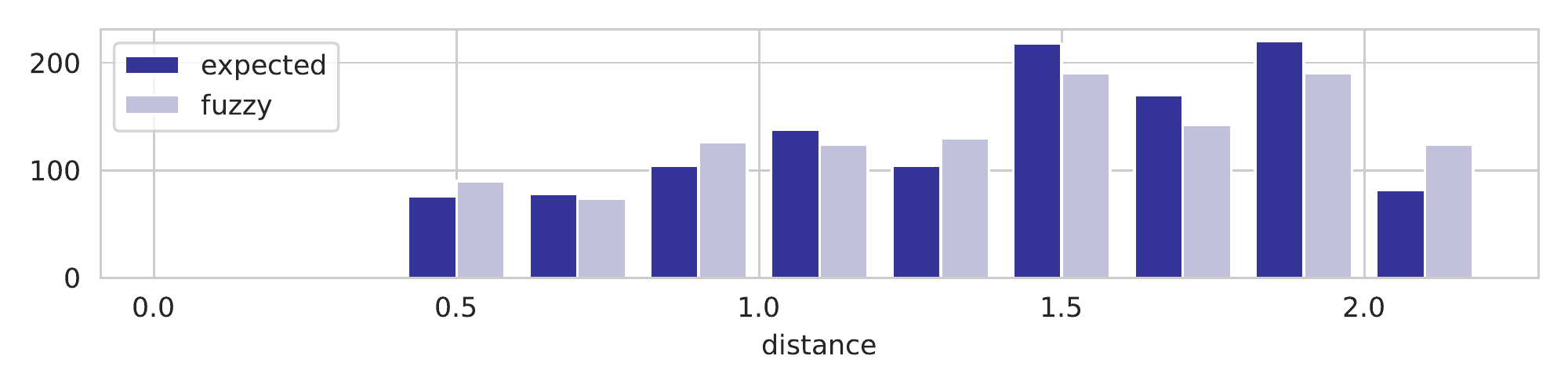}}
    \caption{$c=0.9$ $N=35$}
    \end{subfigure}

  \begin{subfigure}{0.85\textwidth}
{\includegraphics[width=\textwidth]{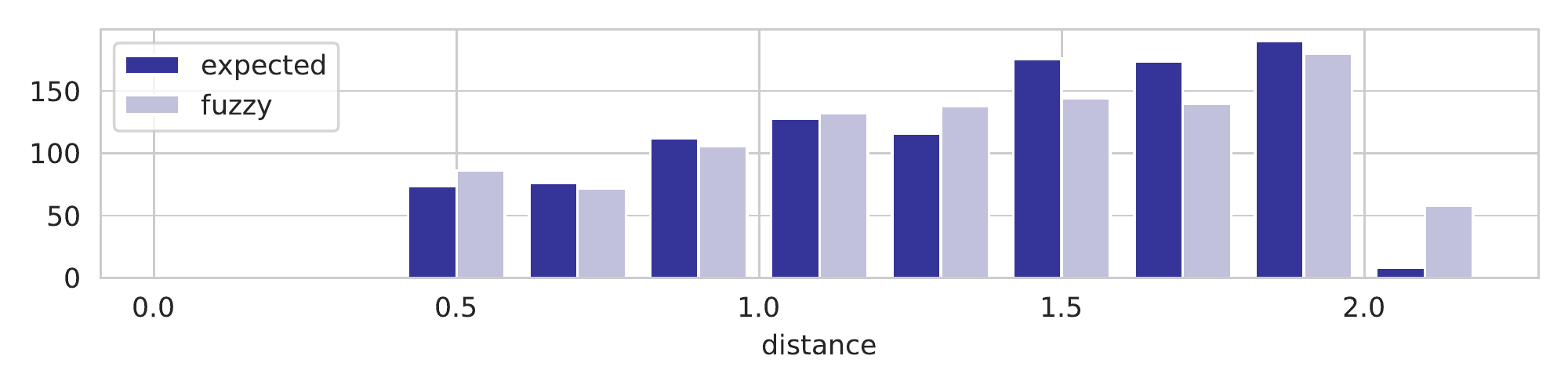}}
\caption{$c=1$ $N=33$}
\end{subfigure}
\caption{Histogram of the distances for $n=8$, for $c=0.5,0.9,1.0$.  The histograms generated from spheroids are dark blue, while the histogram for the fuzzy spaces is light blue. The number of states $N$ is given for each, since we generated as many points for each expected state as we found in the corresponding simulation.}\label{fig:disthistclt1}
\end{figure}
\begin{figure}
  \centering
\begin{subfigure}{0.85\textwidth}
{\includegraphics[width=\textwidth]{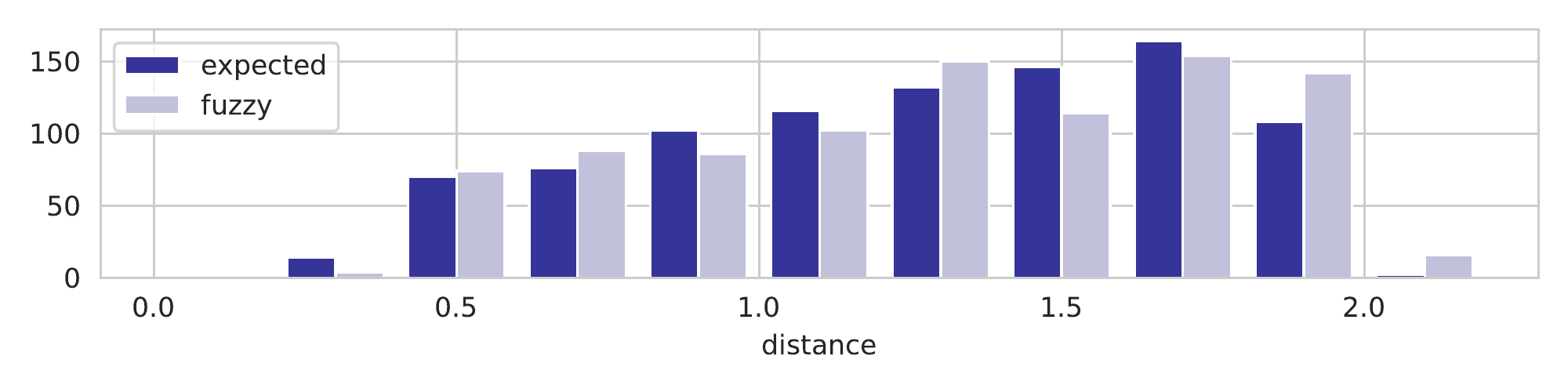}}
\caption{$c=1.1$ $N=31$}
\end{subfigure}

\begin{subfigure}{0.85\textwidth}
{\includegraphics[width=\textwidth]{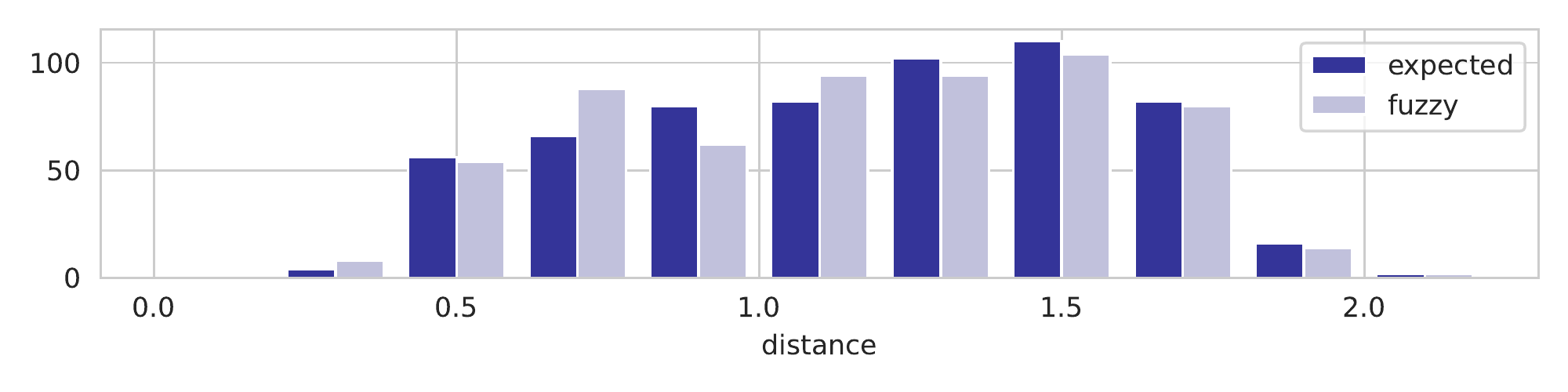}}
\caption{$c=1.5$ $N=25$}
\end{subfigure}

\begin{subfigure}{0.85\textwidth}
{\includegraphics[width=\textwidth]{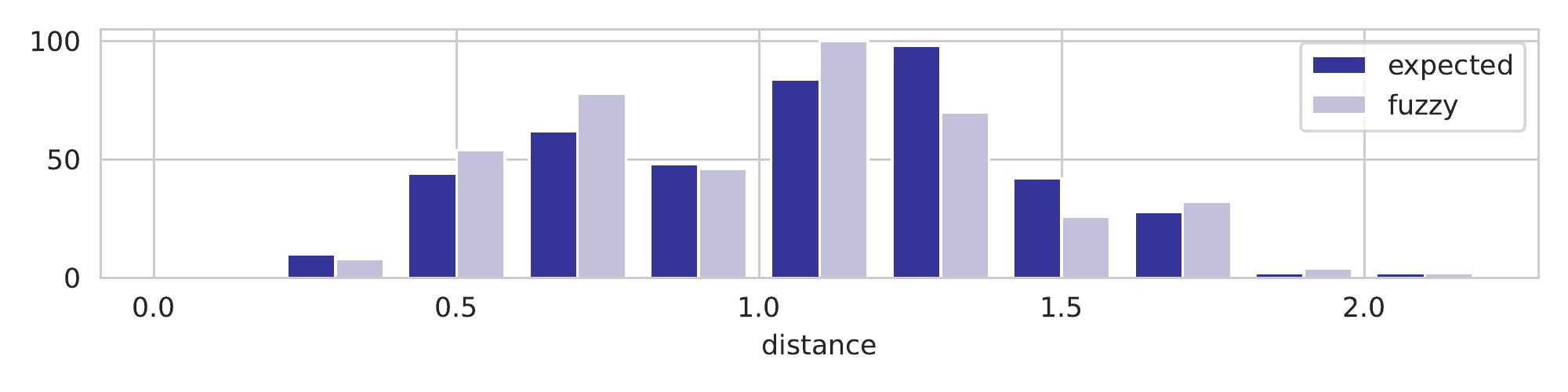}}
\caption{$c=2$ $N=21$}
\end{subfigure}

\begin{subfigure}{0.85\textwidth}
{\includegraphics[width=\textwidth]{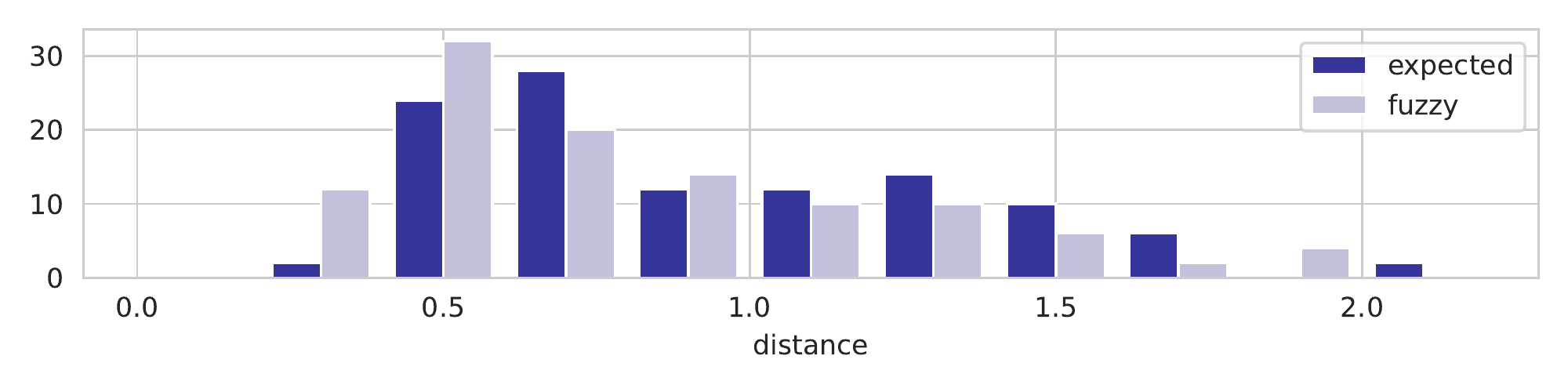}}
\caption{$c=5$ $N=11$}
\end{subfigure}
\caption{Histogram of the distances for $n=8$,  for $c=1.1,1.5,2.0,5.0$.  The  histograms generated from spheroids are dark blue, while the histogram for the fuzzy spaces is light blue. The number of states $N$ is given for each, since we generated as many points for each expected state as we found in the corresponding simulation.}\label{fig:disthistcgt1}
\end{figure}

The results are shown in Figures~\ref{fig:disthistclt1} and~\ref{fig:disthistcgt1} with the data for the fuzzy space embeddings in light blue and the data generated from a continuum embedding in dark blue.
The histograms for the two samples agree well, within the limitation of the small sample size.
Since generating each embedding takes several days it is not practical to generate more for sampling purposes.

We thus conclude that this observable is also well compatible with the expectation that our deformed fuzzy space is an ellipsoid.

\subsection{A more general deformed sphere}
While we do not have an analytic equation for the eigenvalues of the Dirac operator with general deformations, our code can easily deal with arbitrary values of the $c_{ij}$\footnote{$c_0$ is fixed to $1$ however an overall rescaling of $D$ can always set this to $1$.}. 

\begin{figure}
  \begin{subfigure}{0.49\textwidth}
{\includegraphics[width=\textwidth]{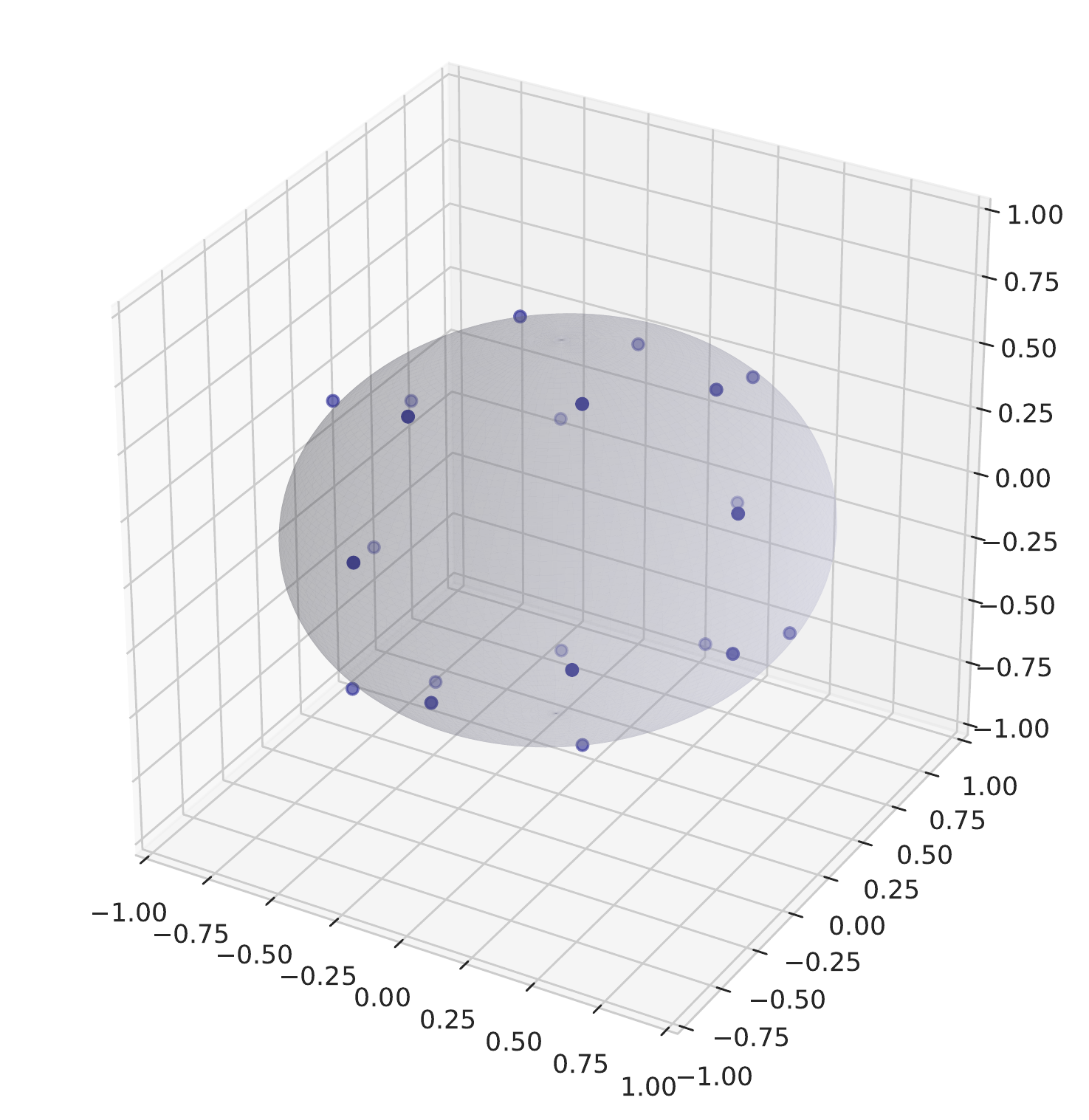}}
\caption{$c_{ij}=(1.1,1.1,1.5)$}
\end{subfigure}
\begin{subfigure}{0.49\textwidth}
{\includegraphics[width=\textwidth]{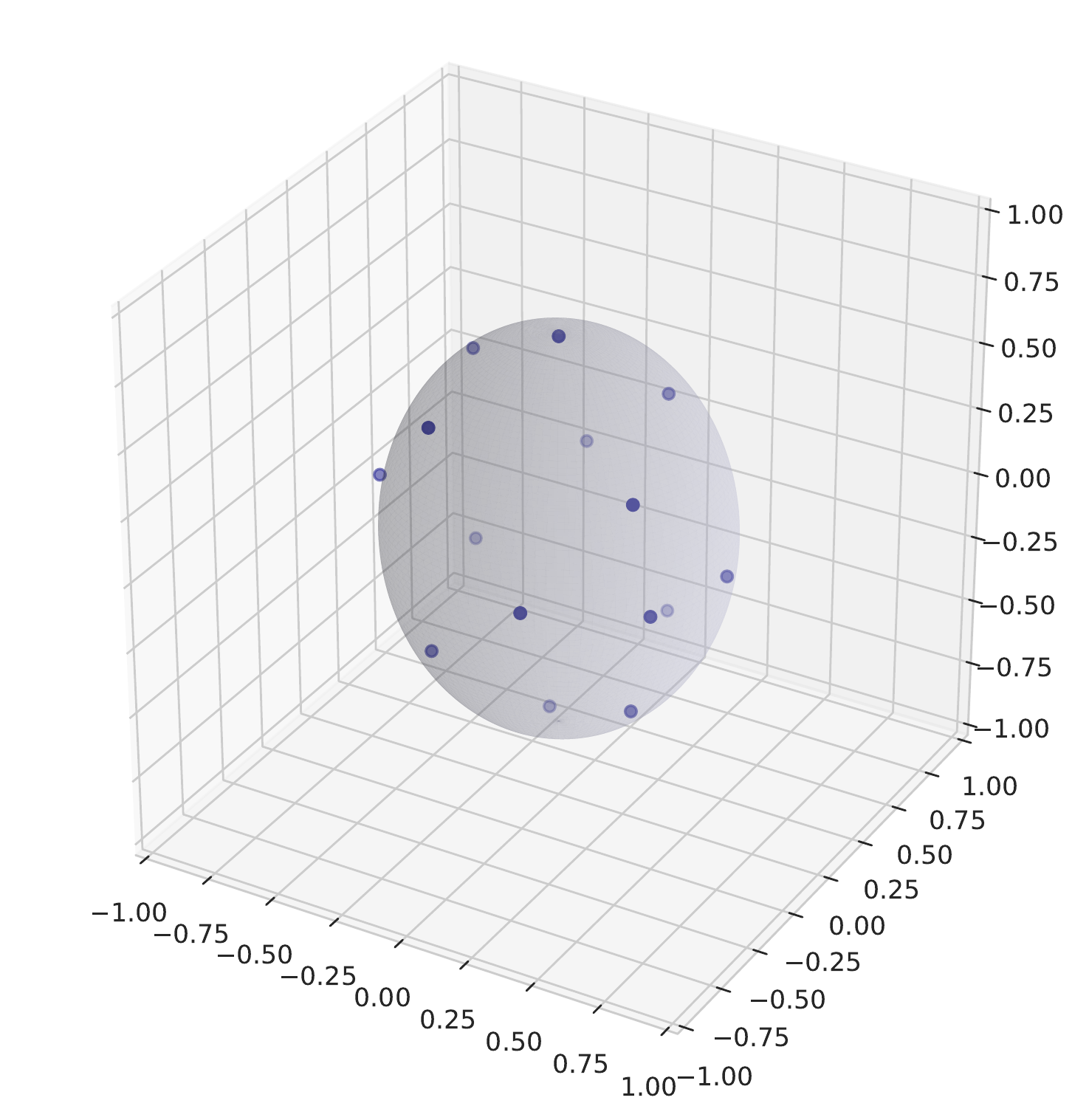}}
\caption{$c_{ij}=(1.1,2.0,1.5)$}
\end{subfigure}

\caption{Two examples of more general deformed spheres.}\label{fig:general_ellipse}
\end{figure}

We have thus tested this, again for $n=8$ and fit the data with ellipsoids as explained above.
In Figure~\ref{fig:general_ellipse} we show this for the two cases of $c_{ij}=(1.1,1.1,1.5)$ and $c_{ij}=(1.1,2.0,1.5)$.
The fit data for all states generated is given in the appendix in table~\ref{tab:allthefits}.
While these fits overall work well, there are several outliers where the algorithm was not able to find a good fit.
This is likely partly due to fact that the deformation leads to smaller volumes, and thus fewer states, which means less data to constrain the fit.
Running a less constrained fit on the computer leaves more opportunities to find fits that minimize the function, but are not actually the `true' hidden geometry.

\begin{figure}
  \centering
  \includegraphics[width=0.7\textwidth]{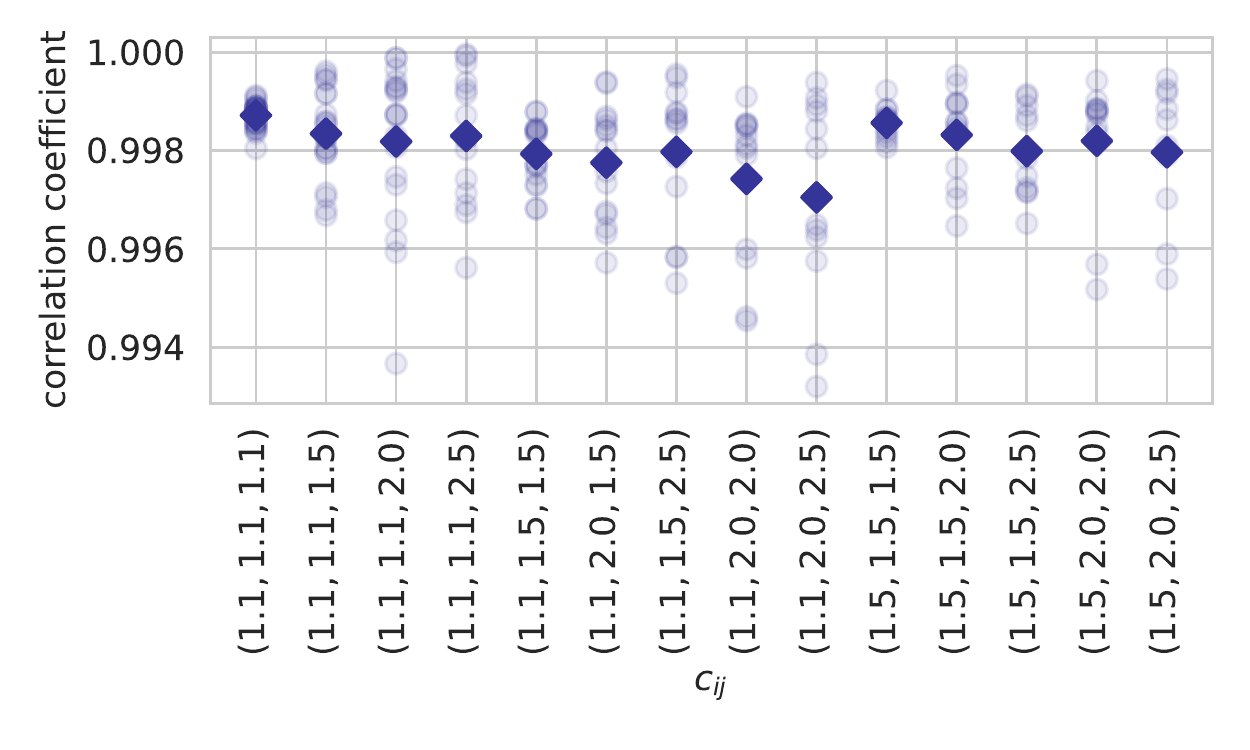}
  \caption{Correlation coefficient for the fuzzy sphere when it is deformed in several directions.}\label{fig:cc_gendef}
\end{figure}

So while our analytic understanding of the eigenvalues is restricted to the simpler case, our code can generate embeddings for the more complicated class of deformed fuzzy spheres, and they can still be fit through ellipsoids.
It thus seems that our qualitative understanding of the geometry works well.
In Figure~\ref{fig:cc_gendef} we show the correlation coefficient between the distances of the points in the embedding and those found using the spectral distances for the multiply deformed sphere.
As for the case with deformation in only a single direction the correlation coefficient is extremely close to $1$, leading us to believe that the multi deformed sphere also works well.

\section{Conclusion and outlook}
In this article we have studied the Dirac operator for a deformed fuzzy sphere.
We found that, simply rescaling the terms in the Dirac operator for the fuzzy sphere does lead to a geometry that can be described as an ellipsoid.
We studied some spectral observables for the deformed fuzzy sphere, among them examining the spectral variance of the object.

Next we used this new geometry as a test case in generalizing the algorithm proposed in~\cite{Glaser_Stern_2021} from truncated spectral triples to finite spectral triples, sometimes also called fuzzy spaces.
We were able to obtain visualizations of the deformed fuzzy sphere, which agreed well with the expectation.
Fitting the length of the axes of the embedded geometries led to values close to the expected values, within the uncertainties arising from single instance numerical work.

Another method to test the veracity of the embedding was to study the correlation between the distances as embedded, and the distances calculated.
This measures if the geometries can actually be embedded in flat $3$ dimensional space.
The correlation coefficients all came out extremely close to $1$ confirming that the geometries embed well.

This work is a first step into visualizing finite spectral triples.
A promising next step would be to visualize the fuzzy torus~\cite{Barrett:2019ize}.
One particularly interesting property of the torus in this context is that it can only embed into $2$ or $4$ dimensions, it would thus be fascinating to see how trying to embed it into the `wrong' dimension changes the results, and to compare its embedding in these two cases.

This would of course only be first steps towards a general purpose code to generate images of any finite spectral triple one might write down.
This code would then be very useful in trying to understand the results arising from the research program into random finite spectral triples.

The code used to generate the visualizations of the deformed fuzzy sphere is available on GitHub~\cite{github_code}.
The full set of data generated for this article is also available for download, together with some scripts to start exploring them, in the encyclopedia of quantum geometries~\cite{glaser_l_2023_7864066}.

\section*{Acknowledgements}
I am grateful to Abel Stern, for continuing discussions about this project, even though he has left academia.
I have also had great discussions helping to understand the spectrum with John Barrett and Thomas Laird.
I'd also like to thank the Erwin Schrödinger Institute in Vienna, where John and Thomas were visiting during some of these discussions.
\printbibliography

@article{Verhoeven_2023, title={Geometry in spectral triples: Immersions and fermionic fuzzy geometries}, url={https://ir.lib.uwo.ca/etd/9561}, journal={Electronic Thesis and Dissertation Repository}, author={Verhoeven, Luuk}, year={2023}, month=aug }

@article{Hessam_Khalkhali_Pagliaroli_2023, title={Double scaling limits of Dirac ensembles and Liouville quantum gravity}, volume={56}, DOI={10.1088/1751-8121/accfd6}, number={22}, journal={J. Phys. A}, author={Hessam, Hamed and Khalkhali, Masoud and Pagliaroli, Nathan}, year={2023}, pages={225201} }

@article{Pérez-Sánchez_2021, title={On Multimatrix Models Motivated by Random Noncommutative Geometry I: The Functional Renormalization Group as a Flow in the Free Algebra}, volume={22}, ISSN={1424-0661}, DOI={10.1007/s00023-021-01025-4}, abstractNote={Random noncommutative geometry can be seen as a Euclidean path-integral quantization approach to the theory defined by the Spectral Action in noncommutative geometry (NCG). With the aim of investigating phase transitions in random NCG of arbitrary dimension, we study the nonperturbative Functional Renormalization Group for multimatrix models whose action consists of noncommutative polynomials in Hermitian and anti-Hermitian matrices. Such structure is dictated by the Spectral Action for the Dirac operator in Barrett’s spectral triple formulation of fuzzy spaces. The present mathematically rigorous treatment puts forward “coordinate-free” language that might be useful also elsewhere, all the more so because our approach holds for general multimatrix models. The toolkit is a noncommutative calculus on the free algebra that allows to describe the generator of the renormalization group flow—a noncommutative Laplacian introduced here—in terms of Voiculescu’s cyclic gradient and Rota–Sagan–Stein noncommutative derivative. We explore the algebraic structure of the Functional Renormalization Group equation and, as an application of this formalism, we find the $$beta $$-functions, identify the fixed points in the large-N limit and obtain the critical exponents of two-dimensional geometries in two different signatures.}, number={9}, journal={Annales Henri Poincaré}, author={Pérez-Sánchez, Carlos I.}, year={2021}, month=sep, pages={3095–3148}, language={en} }

@article{Barrett_Druce_Glaser_2019, title={Spectral estimators for finite non-commutative geometries}, volume={52}, ISSN={1751-8121}, DOI={10.1088/1751-8121/ab22f8}, abstractNote={A finite non-commutative geometry consists of a fuzzy space together with a Dirac operator satisfying the axioms of a real spectral triple. This paper addresses the question of how to extract information about these geometries from the spectrum of the Dirac operator. Since the Dirac operator is a finite-dimensional matrix, the usual asymptotics of the eigenvalues makes no sense and is replaced by measurements of the spectrum at a finite energy scale. The spectral dimension of the square of the Dirac operator is improved to provide a new spectral measure of the dimension of a space called the spectral variance. Similarly, the volume of a space can be computed from the spectrum once the dimension is known. Two methods of doing this are investigated: the well-known Dixmier trace and a recent improvement due to Abel Stern. Finally, the distance between two geometries is investigated by comparing the spectral zeta functions using the method of Cornelissen and Kontogeorgis. All of these techniques are tested on the explicit examples of the fuzzy spheres and fuzzy tori, which can be regarded as approximations of the usual Riemannian sphere and flat tori. Then they are applied to characterise some random fuzzy spaces using data generated by a Monte Carlo simulation.}, number={27}, journal={Journal of Physics A: Mathematical and Theoretical}, author={Barrett, John W. and Druce, Paul and Glaser, L}, year={2019}, month={Jun}, pages={275203}, language={en} }

@article{future, title={Work in progress}, author={Barrett, John W. and Glaser, L}}

@article{Barrett_2015, title={Matrix geometries and fuzzy spaces as finite spectral triples}, volume={56}, ISSN={0022-2488, 1089-7658}, DOI={10.1063/1.4927224}, abstractNote={A class of real spectral triples that are similar in structure to a Riemannian manifold but have a finite-dimensional Hilbert space is defined and investigated, determining a general form for the Dirac operator. Examples include fuzzy spaces defined as real spectral triples. Fuzzy 2-spheres are investigated in detail, and it is shown that the fuzzy analogues correspond to two spinor fields on the commutative sphere. In some cases, it is necessary to add a mass mixing matrix to the commutative Dirac operator to get a precise agreement for the eigenvalues.}, number={8}, journal={Journal of Mathematical Physics}, author={Barrett, John W.}, year={2015}, month={Aug}, pages={082301} }

@article{Glaser_Stern_2021, title={Reconstructing manifolds from truncations of spectral triples}, volume={159}, ISSN={0393-0440}, DOI={https://doi.org/10.1016/j.geomphys.2020.103921}, abstractNote={We explore the geometric implications of introducing a spectral cut-off on compact Riemannian manifolds. This is naturally phrased in the framework of non-commutative geometry, where we work with spectral triples that are truncated by spectral projections of Dirac-type operators. We associate a metric space of ‘localized’ states to each truncation. The Gromov–Hausdorff limit of these spaces is then shown to equal the underlying manifold one started with. This leads us to propose a computational algorithm that allows us to approximate these metric spaces from the finite-dimensional truncated spectral data. We subsequently develop a technique for embedding the resulting metric graphs in Euclidean space to asymptotically recover an isometric embedding of the limit. We test these algorithms on the truncation of sphere and a recently investigated perturbation thereof.}, journal={Journal of Geometry and Physics}, author={Glaser, L and Stern, Abel B.}, year={2021}, pages={103921} }

@article{Ambjørn_Jurkiewicz_Loll_2005, title={The Spectral Dimension of the Universe is Scale Dependent}, volume={95}, DOI={10.1103/PhysRevLett.95.171301}, abstractNote={We measure the spectral dimension of universes emerging from nonperturbative quantum gravity, defined through state sums of causal triangulated geometries. While four dimensional on large scales, the quantum universe appears two dimensional at short distances. We conclude that quantum gravity may be “self-renormalizing” at the Planck scale, by virtue of a mechanism of dynamical dimensional reduction.}, number={17}, journal={Physical Review Letters}, author={Ambjørn, J. and Jurkiewicz, J. and Loll, R.}, year={2005}, month={Oct}, pages={171301} }

@article{Chamseddine_Connes_Mukhanov_2014, title={Geometry and the Quantum: Basics}, volume={2014}, ISSN={1029-8479}, url={http://arxiv.org/abs/1411.0977}, DOI={10.1007/JHEP12(2014)098}, abstractNote={Motivated by the construction of spectral manifolds in noncommutative geometry, we introduce a higher degree Heisenberg commutation relation involving the Dirac operator and the Feynman slash of scalar fields. This commutation relation appears in two versions, one sided and two sided. It implies the quantization of the volume. In the one-sided case it implies that the manifold decomposes into a disconnected sum of spheres which will represent quanta of geometry. The two sided version in dimension 4 predicts the two algebras M_2(H) and M_4(C) which are the algebraic constituents of the Standard Model of particle physics. This taken together with the non-commutative algebra of functions allows one to reconstruct, using the spectral action, the Lagrangian of gravity coupled with the Standard Model. We show that any connected Riemannian Spin 4-manifold with quantized volume >4 (in suitable units) appears as an irreducible representation of the two-sided commutation relations in dimension 4 and that these representations give a seductive model of the “particle picture” for a theory of quantum gravity in which both the Einstein geometric standpoint and the Standard Model emerge from Quantum Mechanics. Physical applications of this quantization scheme will follow in a separate publication.}, note={arXiv: 1411.0977}, number={12}, journal={Journal of High Energy Physics}, author={Chamseddine, Ali H. and Connes, Alain and Mukhanov, Viatcheslav}, year={2014}, month={Dec} }

@article{Schneiderbauer_Steinacker_2016, title={Measuring finite Quantum Geometries via Quasi-Coherent States}, volume={49}, ISSN={1751-8113, 1751-8121}, DOI={10.1088/1751-8113/49/28/285301}, abstractNote={We develop a systematic approach to determine and measure numerically the geometry of generic quantum or “fuzzy” geometries realized by a set of finite-dimensional hermitian matrices. The method is designed to recover the semi-classical limit of quantized symplectic spaces embedded in $mathbb{R}^d$ including the well-known examples of fuzzy spaces, but it applies much more generally. The central tool is provided by quasi-coherent states, which are defined as ground states of Laplace- or Dirac operators corresponding to localized point branes in target space. The displacement energy of these quasi-coherent states is used to extract the local dimension and tangent space of the semi-classical geometry, and provides a measure for the quality and self-consistency of the semi-classical approximation. The method is discussed and tested with various examples, and implemented in an open-source Mathematica package.}, note={arXiv: 1601.08007}, number={28}, journal={Journal of Physics A: Mathematical and Theoretical}, author={Schneiderbauer, Lukas and Steinacker, Harold C.}, year={2016}, month={Jul}, pages={285301} }

@article{Barrett_Glaser_2016, title={Monte Carlo simulations of random non-commutative geometries}, volume={A49}, DOI={10.1088/1751-8113/49/24/245001}, abstractNote={Random non-commutative geometries are introduced by integrating over the space of Dirac operators that form a spectral triple with a fixed algebra and Hilbert space. The cases with the simplest types of Clifford algebra are investigated using Monte Carlo simulations to compute the integrals. Various qualitatively different types of behaviour of these random Dirac operators are exhibited. Some features are explained in terms of the theory of random matrices but other phenomena remain mysterious. Some of the models with a quartic action of symmetry-breaking type display a phase transition. Close to the phase transition the spectrum of a typical Dirac operator shows manifold-like behaviour for the eigenvalues below a cut-off scale.}, journal={J.Phys.}, author={Barrett, John W. and Glaser, Lisa}, year={2016}, month={May}, pages={245001} }

@book{Connes_1994, address={San Diego}, title={Noncommutative Geometry}, ISBN={978-0-12-185860-5}, abstractNote={This English version of the path-breaking French book on this subject gives the definitive treatment of the revolutionary approach to measure theory, geometry, and mathematical physics developed by Alain Connes. Profusely illustrated and invitingly written, this book is ideal for anyone who wants to know what noncommutative geometry is, what it can do, or how it can be used in various areas of mathematics, quantization, and elementary particles and fields. It includes features such as: first full treatment of the subject and its applications; written by the pioneer of this field; broad applications in mathematics; of interest across most fields; ideal as an introduction and survey; examples treated include: @subbul; the space of Penrose tilings; the space of leaves of a foliation; the space of irreducible unitary representations of a discrete group; the phase space in quantum mechanics; the Brillouin zone in the quantum Hall effect; and a model of space time.}, publisher={Academic Press}, author={Connes, Alain}, year={1994}, month={Dec}, language={English} }

@ARTICLE{2020SciPy-NMeth,
  author  = {Virtanen, Pauli and Gommers, Ralf and Oliphant, Travis E. and
            Haberland, Matt and Reddy, Tyler and Cournapeau, David and
            Burovski, Evgeni and Peterson, Pearu and Weckesser, Warren and
            Bright, Jonathan and {van der Walt}, St{\'e}fan J. and
            Brett, Matthew and Wilson, Joshua and Millman, K. Jarrod and
            Mayorov, Nikolay and Nelson, Andrew R. J. and Jones, Eric and
            Kern, Robert and Larson, Eric and Carey, C J and
            Polat, {\.I}lhan and Feng, Yu and Moore, Eric W. and
            {VanderPlas}, Jake and Laxalde, Denis and Perktold, Josef and
            Cimrman, Robert and Henriksen, Ian and Quintero, E. A. and
            Harris, Charles R. and Archibald, Anne M. and
            Ribeiro, Ant{\^o}nio H. and Pedregosa, Fabian and
            {van Mulbregt}, Paul and {SciPy 1.0 Contributors}},
  title   = {{{SciPy} 1.0: Fundamental Algorithms for Scientific
            Computing in Python}},
  journal = {Nature Methods},
  year    = {2020},
  volume  = {17},
  pages   = {261--272},
  adsurl  = {https://rdcu.be/b08Wh},
  doi     = {10.1038/s41592-019-0686-2},
}

@article{Stern_2018, title={Finite-rank approximations of spectral zeta residues}, ISSN={1573-0530}, url={https://doi.org/10.1007/s11005-018-1117-5}, DOI={10.1007/s11005-018-1117-5}, abstractNote={We use the asymptotic expansion of the heat trace to express all residues of spectral zeta functions as regularized sums over the spectrum. The method extends to those spectral zeta functions that are localized by a bounded operator.}, journal={Letters in Mathematical Physics}, author={Stern, Abel B.}, year={2018}, month={Jul}, language={en} }

@article{Barrett:2019ize,
    author = "Barrett, John W. and Gaunt, James",
    title = "{Finite spectral triples for the fuzzy torus}",
    eprint = "1908.06796",
    archivePrefix = "arXiv",
    primaryClass = "math.QA",
    month = "8",
    year = "2019"
}

@misc{github_code,
  author = {L Glaser},
  title = {Deformed fuzzy sphere visualisation},
  year = {2023},
  publisher = {GitHub},
  journal = {GitHub repository},
  howpublished = {\url{https://github.com/LisaGlaser/deformed_sphere_visualisation}}
  }

@dataset{glaser_l_2023_7864066,
  author       = {Glaser, L},
  title        = {Deformed fuzzy spheres},
  month        = apr,
  year         = 2023,
  publisher    = {Zenodo},
  version      = {v2},
  doi          = {10.5281/zenodo.7864066},
  url          = {https://doi.org/10.5281/zenodo.7864066}
}

@phdthesis{MauroDArcangelo_2022, address={Nottingham}, title={Numerical Simulation of Random Dirac Operators}, school={University of Nottingham}, author={Mauro D’Arcangelo}, year={2022} }

@article{Azarfar_Khalkhali_2019, title={Random Finite Noncommutative Geometries and Topological Recursion}, url={http://arxiv.org/abs/1906.09362}, abstractNote={In this paper we investigate a model for quantum gravity on finite noncommutative spaces using the theory of blobbed topological recursion. The model is based on a particular class of random finite real spectral triples ${(mathcal{A}, mathcal{H}, D , gamma , J) ,}$, called random matrix geometries of type ${(1,0) ,}$, with a fixed fermion space ${(mathcal{A}, mathcal{H}, gamma , J) ,}$, and a distribution of the form ${e^{- mathcal{S} (D)} {mathop{}!mathrm{d}} D}$ over the moduli space of Dirac operators. The action functional ${mathcal{S} (D)}$ is considered to be a sum of terms of the form ${prod_{i=1}^s mathrm{Tr} left( {D^{n_i}} right)}$ for arbitrary ${s geqslant 1 ,}$. The Schwinger-Dyson equations satisfied by the connected correlators ${W_n}$ of the corresponding multi-trace formal 1-Hermitian matrix model are derived by a differential geometric approach. It is shown that the coefficients ${W_{g,n}}$ of the large $N$ expansion of ${W_n}$’s enumerate discrete surfaces, called stuffed maps, whose building blocks are of particular topologies. The spectral curve ${left( {Sigma , omega_{0,1} , omega_{0,2}} right)}$ of the model is investigated in detail. In particular, we derive an explicit expression for the fundamental symmetric bidifferential ${omega_{0,2}}$ in terms of the formal parameters of the model.}, note={arXiv: 1906.09362}, journal={arXiv:1906.09362 [hep-th, physics:math-ph]}, author={Azarfar, Shahab and Khalkhali, Masoud}, year={2019}, month={Jun} }

@article{Khalkhali_Pagliaroli_2020, title={Phase Transition in Random Noncommutative Geometries}, url={http://arxiv.org/abs/2006.02891}, DOI={10.1088/1751-8121/abd190}, abstractNote={We present an analytic proof of the existence of phase transition in the large $N$ limit of certain random noncommutaitve geometries. These geometries can be expressed as ensembles of Dirac operators. When they reduce to single matrix ensembles, one can apply the Coulomb gas method to find the empirical spectral distribution. We elaborate on the nature of the large $N$ spectral distribution of the Dirac operator itself. Furthermore, we show that these models exhibit both a single and double cut region for certain values of the order parameter and find the exact value where the transition occurs.}, note={arXiv: 2006.02891}, journal={arXiv:2006.02891 [hep-th, physics:math-ph]}, author={Khalkhali, Masoud and Pagliaroli, Nathan}, year={2020}, month={Dec} }

@article{Khalkhali_Pagliaroli_2021, title={Spectral Statistics of Dirac Ensembles}, url={http://arxiv.org/abs/2109.12741}, abstractNote={In this paper we find spectral properties in the large $N$ limit of Dirac operators that come from random finite noncommutative geometries. In particular for a Gaussian potential the limiting eigenvalue spectrum is shown to be universal regardless of the geometry and is given by the convolution of the semicircle law with itself. For simple non-Gaussian models this convolution property is also evident. In order to prove these results we show that a wide class of multi-trace multimatrix models have a genus expansion.}, note={arXiv: 2109.12741}, journal={arXiv:2109.12741 [hep-th, physics:math-ph]}, author={Khalkhali, Masoud and Pagliaroli, Nathan}, year={2021}, month={Sep} }

@article{Hessam_Khalkhali_Pagliaroli_2021, title={Bootstrapping Dirac Ensembles}, url={http://arxiv.org/abs/2107.10333}, abstractNote={We apply the bootstrap technique to find the moments of certain multi-trace and multi-matrix random matrix models suggested by noncommutative geometry. Using bootstrapping we are able to find the relationships between the coupling constant of these models and their second moments. Using the Schwinger-Dyson equations, all other moments can be expressed in terms of the coupling constant and the second moment. Explicit relations for higher moments are obtained.}, note={arXiv: 2107.10333}, journal={arXiv:2107.10333 [hep-th, physics:math-ph]}, author={Hessam, Hamed and Khalkhali, Masoud and Pagliaroli, Nathan}, year={2021}, month={Jul} }

@article{Glaser_2017, title={Scaling behaviour in random non-commutative geometries}, volume={50}, ISSN={1751-8121}, DOI={10.1088/1751-8121/aa7424}, abstractNote={Random non-commutative geometries are a novel approach to taking a non-perturbative path integral over geometries. They were introduced in Barrett and Glaser (2015 arXiv:1510.01377), where a first examination was performed. During this examination we found that some geometries show indications of a phase transition. In this article we explore this phase transition further for geometries of type (1, 1), (2, 0), and (1, 3). We determine the pseudo-critical points of these geometries and explore how some of the observables scale with the system size. We also undertake first steps towards understanding the critical behaviour through correlations and in determining critical exponents of the system.}, number={27}, journal={Journal of Physics A: Mathematical and Theoretical}, author={Glaser, L}, year={2017}, pages={275201}, language={en} }

@article{Grosse_Prešnajder, title={The dirac operator on the fuzzy sphere}, volume={33}, ISSN={0377-9017, 1573-0530}, DOI={10.1007/BF00739805}, abstractNote={We introduce the fuzzy analog of spinor bundles over the sphere on which the noncommutative analog of the Dirac operator acts. We construct the complete set of eigenstates including zero modes. In the commutative limit, we recover known results.}, number={2}, journal={Letters in Mathematical Physics}, author={Grosse, H. and Prešnajder, P.}, pages={171–181}, language={en} }
\begin{appendix}
\section{General deformed sphere}
\begin{table}[h]
  \caption{Fit data for general deformed spheres.}\label{tab:allthefits}
  \small

  \begin{tabular}{r r  r r  r}
    \toprule
   $c_{ij}$ &  expected axes & best fit axes & angle of axis & $f$/ d.o.f. \\  \midrule 
   $ (1.10,1.10,1.10)$ & $ (0.83,0.83,0.83)$ &$ (0.89,0.90,0.91)$ & $ (0.60,0.00)$ &$ 0.0036  $  \\ 
   $ (1.10,1.10,1.50)$ & $ (0.61,0.61,0.83)$ &$ (0.72,0.72,0.93)$ & $ (0.01,0.70)$ &$ 0.0013  $  \\ 
   $ (1.10,1.10,2.00)$ & $ (0.45,0.45,0.83)$ &$ (0.56,0.57,0.97)$ & $ (0.34,-0.00)$ &$ 0.0087  $  \\ 
   $ (1.10,1.10,2.50)$ & $ (0.36,0.36,0.83)$ &$ (0.47,0.59,0.96)$ & $ (0.00,0.00)$ &$ 0.1137  $  \\ 
   $ (1.10,1.50,1.50)$ & $ (0.44,0.61,0.61)$ &$ (0.67,0.73,0.73)$ & $ (-0.00,0.71)$ &$ 0.0024  $  \\ 
   $ (1.10,1.50,2.50)$ & $ (0.27,0.36,0.61)$ &$ (0.44,0.49,0.76)$ & $ (0.82,1.58)$ &$ 0.0315  $  \\ 
   $ (1.10,1.50,2.00)$ & $ (0.33,0.45,0.61)$ &$ (0.54,0.57,0.77)$ & $ (0.40,0.38)$ &$ 0.0125  $  \\ 
   $ (1.10,2.00,2.00)$ & $ (0.25,0.45,0.45)$ &$ (0.49,0.58,0.59)$ & $ (0.07,-0.00)$ &$ 0.0101  $  \\ 
   $ (1.10,2.00,2.50)$ & $ (0.20,0.36,0.45)$ &$ (0.43,0.48,0.58)$ & $ (0.89,1.64)$ &$ 0.0249  $  \\ 
   $ (1.50,1.50,1.50)$ & $ (0.44,0.44,0.44)$ &$ (0.66,0.66,0.66)$ & $ (0.01,0.00)$ &$ 0.0090  $  \\ 
   $ (1.50,1.50,2.00)$ & $ (0.33,0.33,0.44)$ &$ (0.54,0.54,0.68)$ & $ (0.25,1.55)$ &$ 0.0144  $  \\ 
   $ (1.50,1.50,2.50)$ & $ (0.27,0.27,0.44)$ &$ (0.46,0.47,0.64)$ & $ (0.11,0.16)$ &$ 0.0229  $  \\ 
   $ (1.50,2.00,2.00)$ & $ (0.25,0.33,0.33)$ &$ (0.49,0.55,0.56)$ & $ (0.00,0.30)$ &$ 0.0342  $  \\ 
   $ (1.50,2.00,2.50)$ & $ (0.20,0.27,0.33)$ &$ (0.43,0.52,0.53)$ & $ (0.00,0.00)$ &$ 0.0458  $  \\   
 \bottomrule
  \end{tabular}
  
  \end{table}
  
\end{appendix}

\end{document}